\documentclass[review]{elsarticle}

\usepackage{bbm} 
\usepackage{caption}
\usepackage{subcaption}
\usepackage{color}
\usepackage{multirow}
\usepackage{textcomp}
\usepackage{hyperref}
\usepackage{color}
\usepackage{amsmath}
\usepackage{amssymb}
\usepackage{graphicx}
\usepackage{hyperref} 
\usepackage{xurl} 
\hypersetup{
    colorlinks=true,
    linkcolor=blue,
    filecolor=magenta,      
    urlcolor=cyan,
    citecolor=green,
}

\journal{arXiv}

\newtheorem{assumption}{Assumption}
\newtheorem{remark}{Remark}
\newtheorem{lemma}{Lemma}

\usepackage[british]{babel} 
\begin{document}

\begin{frontmatter}

\title{Active Disturbance Rejection Control (ADRC) Toolbox\\ for MATLAB/Simulink}

\author{Krzysztof Łakomy}
\ead{krzysztof.lakomy92@gmail.com}

\author[PUT]{Wojciech Giernacki}
\ead{wojciech.giernacki@put.poznan.pl}

\author[PUT]{Jacek Michalski}
\ead{jacek.michalski@put.poznan.pl}

\author[JNU]{Rafal Madonski\corref{mycorrespondingauthor}}
\ead{rafal.madonski@jnu.edu.cn}
\ead{rafal.madonski@gmail.com}

\cortext[mycorrespondingauthor]{Corresponding author}

\address[PUT]{
Institute of Robotics and Machine Intelligence, Faculty of Control, Robotics and Electrical Engineering, Poznan University of Technology, Piotrowo 3a, 60-965 Poznan, Poland}
\address[JNU]{
Energy and Electricity Research Center, Jinan University, 519070 Zhuhai, P. R. China}

\begin{abstract}
In this study, an active disturbance rejection control (ADRC) toolbox for MATLAB\slash Simulink is introduced. Although ADRC has already been established as a powerful robust control framework with successful industrial implementations and strong theoretical foundations, a comprehensive tool for computer-aided design of ADRC has not been developed until now. The proposed open-source \textit{ADRC Toolbox} is a response to the growing need in the scientific community and the control industry for a straightforward software application of the ADRC methodology. Its main purpose is to fill the gap between the current theories and applications of ADRC and to provide an easy-to-use solution for users in various control fields who want to employ the ADRC scheme in their applications. The ADRC Toolbox contains a single, general-purpose, drag-and-drop function block that allows the synthesis of a predefined ADRC-based strategy with minimal design effort. \textcolor{black}{Additionally, its open structure allows creation of custom control solutions. The efficacy of the ADRC Toolbox is validated through both simulations and hardware experiments, which were conducted using a variety of problems known in the motion, process, and power control areas}\footnote{The contents of this paper refer to version 1.1.3 of the ADRC Toolbox; the current version and the repository of previous releases are available at \url{https://www.mathworks.com/matlabcentral/fileexchange/102249-active-disturbance-rejection-control-adrc-toolbox}}.
\end{abstract}



\end{frontmatter}

\section{Introduction}
\label{sec:Intro}

With the development of modern control techniques, new reliable tools for rapid prototyping are being developed. The intensive expansion of the MATLAB\slash Simulink\footnote{MATLAB and Simulink are registered trademarks of MathWorks, Inc.} software environment over the past decade in both academia and industry strongly supports the current trend of using high-level programming for designing, analysing, and applying control systems. By analysing the market and scientific literature, one can find various software toolboxes that support the synthesis of control systems, with emphasis on identifying dynamic systems \cite{CEPUNIT}, modelling and solving optimisation problems \cite{YALMIPtoolbox}, signal differentiators \cite{REDtoolbox}, disturbance observers \cite{DOBtoolbox}, fractional-order control~\cite{FOMCONtoolbox}, aerospace vehicle applications~\cite{AeroToolbox}, or rapid prototyping~\cite{Momir}.


Several types of supporting software solutions have been developed for control systems based on active disturbance rejection control (ADRC) methodology. ADRC is a highly practical control framework introduced to the general audience in~\cite{HanAdrc2001,HanTIE}\footnote{The ADRC was formally proposed in 1998 by Prof. Jingqing Han and initial papers are in Chinese (see~\cite{HanAdrc2001} and references therein).} and later streamlined by \cite{GaoPolePlacement}. Since its inception, it has attracted considerable attention from both scholars and industry practitioners. The effectiveness of its disturbance-centric approach (\cite{GaoCentrality}) for handling uncertainties in governed systems has been verified  \textcolor{black}{through both theoretical analyses~\cite{ADRCanalLyapunov} and} numerous control problems in motion, process, and power control areas~\cite{WenchaoOverview,RecentAppADRC,HeTingCEP}. The combination of modern control elements with the pragmatism of the minimum-modelling approach has made ADRC an interesting alternative to proportional integral derivative (PID) controllers~\cite{HerbstTransfer}. Recent surveys \cite{DOB30years,DOBCoverview,MadoSurvey} have captured the significant impact that ADRC has had on the control field in the last two decades. Conquering market share from popular PID-type controllers, however, is not an easy feat and demands improvements in many areas. Many studies have been conducted thus far (see a summary in~\cite{HerbstFootprint}), and it is thus not surprising that various software programs have been developed over the years to facilitate the implementation of ADRC. 


\textcolor{black}{The ADRC-based systems was implemented so far in a variety of programming languages, like C~\cite{AddOnModule} or Python~\cite{kicki2021tuning}, and in different programming environments like the an open-source Scilab/Xcos~\cite{ADRCtoolboxScilab} and OpenPCS~\cite{TimeDelayZhao}.} In \cite{ADRCtoolboxPLC3,ADRCtoolboxPLC1,ADRCtoolboxPLC2}, various general-purpose function blocks were developed for the industrial implementation of ADRC in programmable logic controllers (PLCs). Two versions of the ADRC library for PLCs were proposed in~\cite{ADRCtoolboxPLC3}, one as a standard version (with basic functionalities), and the other dedicated to experienced ADRC users allowing for partial dynamic model incorporation, relative disturbance order selection, and nonlinear estimation error function. In~\cite{ADRCtoolboxPLC1,ADRCtoolboxPLC2}, the general purpose ADRC function block addresses practically important aspects of process control, like the impact of the derivative backoff, autotuning, or time delay handling. Although the above Scilab/Xcos and PLC implementations of ADRC are neither dedicated to MATLAB\slash Simulink nor publicly available, they provide important information about parameters and functionalities that a practically appealing ADRC software toolbox should possess.


Regarding software support for ADRC implementation in MATLAB\slash Simulink, to the best of our knowledge, only a few approaches are currently available. Interactive ADRC tuning tools, in the form of graphical user interfaces (GUIs), were developed in~\cite{ADRCGUI1,ADRCGUI2}. In~\cite{ADRCGUI1}, ADRC tuning was realised through a dedicated GUI, where the user inputs both the system model and ADRC parameters. Then, the tool visualises the calculated tracking and disturbance rejection performance (based on some predefined robustness constraints). The tool can also initiate randomised Monte Carlo tests to check robustness against system parametric uncertainty in prescribed ranges. In~\cite{ADRCGUI2}, an interactive ADRC design for flight attitude control was proposed. In particular, computer-aided design software, based on the MATLAB GUI, was designed to integrate the tuning and simulation processes with a stability margin tester.  

A tutorial for ADRC design was published in \cite{Martinez}. The Authors publicly shared their implementation examples on the official \textit{MATLAB File Exchange} website, which is dedicated to free, open-source code sharing. In~\cite{Hu,Jiang,Kia}, MATLAB\slash Simulink module-based ADRC designs were presented. These works describe drag-and-drop creation of different ADRC schemes by manually combining various modules (subsystems) typically seen in a conventional ADRC block diagram, such as the tracking differentiator, extended state observer, and state feedback controller. Although the proposed module-based design offers flexibility in structure selection, and allows the ADRC to be customised to a given control scenario, it inevitably increases the commissioning time and the required know-how because users need to implement and independently tune all the necessary ADRC components separately.

Given the above literature overview of available software solutions for ADRC implementation, a new MATLAB\slash Simulink toolbox is developed in this study. The proposed \textit{ADRC Toolbox} expands the current state-of-the-art solutions by distilling the best practices and most useful features into a single tool for continuous-time implementation of ADRC that provides prospective users with software that:
\begin{itemize}
    \item contains a general-purpose ADRC function block that can be tailored to a given control scenario with minimum tuning parameters;
     \item is a development with functionalities that increase its applicable range and practical appeal;
    \item can be easily incorporated into a programme through drag-and-drop and offers almost plug-and-play operation;
    \item has a minimalistic and user-friendly graphical interface; and
    \item is freely available to the control community as open-source software at \url{https://www.mathworks.com/matlabcentral/fileexchange/102249-active-disturbance-rejection-control-adrc-toolbox}.
\end{itemize}

The goal of the proposed ADRC Toolbox is to provide an intuitive blockset within the MATLAB\slash Simulink environment that will allow users to implement the ADRC structure with relatively little effort and limited knowledge of the details of observer-based control. The introduction of the toolbox aims to decrease the time and effort needed to design and implement basic ADRC schemes, and open new possibilities to swiftly deploy ADRC for real-world control problems.

To validate the efficacy of the proposed ADRC Toolbox for MATLAB\slash Simulink, several tests were conducted utilising selected problems from motion, process, and power control areas that differ from each other in terms of order, type, and structure of the controlled dynamics and the type of acting external disturbance. Some of the validating tests were performed in simulations, whereas others were realised in laboratory hardware. Such diversified groups were deliberately selected to examine the ADRC Toolbox in various control scenarios. \textcolor{black}{The mentioned tests are added in the ADRC Toolbox as examples to allow reproducibility of the results shown in this paper.}



\textbf{Notation.} 
Throughout this paper, $\mathbb{R}$ is used as the set of real numbers and $\mathbb{Z}$ as the set of integers. The $i$-th derivative of the signal $x(t)$ with respect to time $t$ is represented by  $x^{(i)}\triangleq\frac{d^ix(t)}{dt^i}$ for $i\in\mathbb{Z}$, and $\dot{x}\triangleq\frac{dx(t)}{dt}$ and $\ddot{x}\triangleq\frac{d^2x(t)}{dt^2}$ are also used to denote the first and second derivatives. A standardised notation for certain vectors and matrices is used, i.e., $\pmb{A}_n\triangleq\begin{bmatrix} \pmb{0}^{(n-1)\times1} &  \pmb{I}^{(n-1)\times(n-1)} \\ 0 & \pmb{0}^{1\times(n-1)}\end{bmatrix}\in\mathbb{R}^{n\times n}$, $\pmb{b}_n\triangleq[\pmb{0}^{1\times (n-1)} \ 1]^\top\in\mathbb{R}^{n}$, $\pmb{c}_n\triangleq[1 \ \pmb{0}^{1\times(n-1)}]^\top\in\mathbb{R}^{n}$, and $\pmb{d}_n\triangleq[\pmb{0}^{n-2} \ 1 \ 0]^\top\in\mathbb{R}^n$, while $\pmb{0}$ and $\pmb{I}$ correspond, respectively, to zero and identity matrices of appropriate dimensions.
The symbol $\hat{x}(t)$ represents the estimate of the variable $x(t)$, $\textrm{eig}_i(\pmb{A})$ is the $i$-th eigenvalue of matrix $\pmb{A}$, and $\mathbbm{1}(t)$ denotes the unit step function (Heaviside function). Signal $x$, described as $x\sim \mathcal{N}\left(\mu, \sigma^2\right)$, has a normal distribution with standard deviation $\sigma$ and expected value $\mu$. To keep the equations visually appealing, selected functions within the article are written as $f:=f(t)$; hence, direct time dependence is selectively omitted.

\section{Overview of implemented ADRC algorithm}
\label{sec:ADRCpreliminaries}

The general-purpose ADRC function block, developed within the ADRC Toolbox, encapsulates four core concepts and methods: 
\begin{itemize}
    \item \textit{total disturbance}, which results from the aggregation of system uncertainties into a single, matched, observable term;
    \item \textit{extended state}, which results from extending the original model of the controlled dynamics with an auxiliary state variable, representing solely the total disturbance;
    \item \textit{extended state observer}, which is used to reconstruct the total disturbance (and other state variables, if needed) based on available signals and information about the controlled system; and
    \item \color{black} \textit{active disturbance rejection}, which is the process of constantly updating the control signal with the estimated total disturbance, thus allowing on-line compensation of its effects. \color{black}
\end{itemize}

In general, the controller introduced in the ADRC Toolbox (ver. 1.1.3) is designed to achieve control error stabilisation of an $n$-th order nonlinear single-input, single-output (SISO) system, which can be described as
\begin{align}
    \begin{cases}
        x^{(n)}(t) &= f^*\left(x, \dot{x}, \hdots, x^{(n-1)},t\right) + b^*\left(x, \dot{x}, \hdots, x^{(n-1)},t\right)u(t) + d^*(t), \\ 
        y^*(t) &= x(t) + w(t),
    \end{cases}
    \label{eq:originalDynamics}
\end{align}
where $x\in\mathbb{R}$ is the controlled variable, $n\in\mathbb{Z}$ is the order of system dynamics (here, a relative order with respect to the measured signal), $u\in\mathbb{R}$ is the control signal, $d^*\in\mathbb{R}$ represents the external disturbances, and $y^*\in\mathbb{R}$ is the system output perturbed by measurement noise $w\in\mathbb{R}$. Following \cite{lakomy2021isat}, system \eqref{eq:originalDynamics} needs to satisfy the following assumptions for the theoretical analysis (discussed later) to be valid.

\begin{assumption}
    \label{ass:1}
    The variable $x(t)$ and its derivatives up to the $(n-1)$-th order are defined on an arbitrarily large bounded domain $\mathcal{D}_x\triangleq\{[x \ \dot{x} \ \hdots x^{(n-1)}]^\top\in\mathbb{R}:\|[x \ \dot{x} \ \hdots x^{(n-1)}]^\top\|<r_x\}$ for some $r_x>0$.
\end{assumption}

\begin{assumption}
    The measurement noise $w(t)$ belongs to a bounded set  $\mathcal{D}_w\triangleq\{w\in\mathbb{R}:|w|<r_w\}$ for some $r_w>0$.
\end{assumption}

\begin{assumption}
    The external disturbance $d^*(t)$ and its derivative $\dot{d}^*(t)$ belong to the bounded sets $\mathcal{D}_{d^*}\triangleq\{d^*\in\mathbb{R}:d^*<r_{d^*}\}$ and $\mathcal{D}_{\dot{d}^*}\triangleq\{\dot{d}^*\in\mathbb{R}:\dot{d}^*<r_{\dot{d}^*}\}$, for some $r_{d^*},r_{\dot{d}^*}>0$.
\end{assumption}

\begin{assumption}
    \label{ass:4}
    The fields $f^*(x,\dot{x},\hdots,x^{n-1},t):\mathcal{D}_x\times\mathbb{R}\rightarrow\mathbb{R}$ and $b^*(x,\dot{x},\hdots,x^{n-1},t):\mathcal{D}_x\times\mathbb{R}\rightarrow\mathbb{R} / \{0\}$ are continuously differentiable Lipschitz functions.
\end{assumption}

The most common control task in terms of ADRC applications is for $x(t)$ to follow a desired trajectory $x_d(t)\in\mathbb{R}$; in other words, the control task is to stabilise a control error $e(t)\triangleq x_d(t) - x(t)$ in zero. In the case of systems affected by possibly nonlinear external disturbances and measurement noise, one can usually achieve practical stabilisation, that is, one can assure that 
\begin{align}
    \forall_{t\geq T}|e(t)|<\epsilon,
    \label{eq:controlTask}
\end{align}
for some $T>0$ and some arbitrarily small $\epsilon>0$. 

\begin{remark}
    It should be noted that the form of the control task formulated as \eqref{eq:controlTask} is general and can address many conventionally considered control tasks, such as trajectory tracking and set point following. 
\end{remark}

\begin{remark}
\label{rem:linearSystem}
    Although~\eqref{eq:originalDynamics} constitutes a class of nonlinear systems, the control algorithm within the ADRC Toolbox is also applicable to its simplified forms, for example, the group of $n$-th order SISO linear systems.
\end{remark}

Because the control task \eqref{eq:controlTask} is formulated in the control error domain, it will be more convenient to also design the control system for dynamics \eqref{eq:originalDynamics} in the error-domain form, that is,
\begin{align}
\begin{cases}
    e^{(n)}(t) &= x_d^{(n)} - f^*\left(x_d-e, \dot{x}_d - \dot{e}, \hdots, x_d^{(n-1)} - e^{(n-1)},t\right) \\ 
    &- b^*\left(x_d-e, \dot{x}_d - \dot{e}, \hdots, x_d^{(n-1)} - e^{(n-1)},t\right)u(t) - d^*(t), \\ 
    y(t) &= e(t) - w(t),
    \end{cases}
    \label{eq:errorDomainSystem}
\end{align}
where $y\triangleq x_d - y^*\in\mathbb{R}$ is the output of the control error-domain dynamics. 

\begin{assumption}
    \label{ass:5}
    The reference $x_d(t)$ and its derivatives up to the $(n-1)$-th order are defined on an arbitrary large bounded domain $\mathcal{D}_{x_d}\triangleq\{[x_d \ \dot{x}_d \ \hdots x^{(n-1)}_d]^\top\in\mathbb{R}:\|[x_d \ \dot{x}_d \ \hdots x^{(n-1)}_d]^\top\|<r_{x_d}\}$ for some $r_{x_d}>0$.
\end{assumption}


\begin{remark}
    Assumptions~\ref{ass:4} and~\ref{ass:5} can be relaxed based on the results shown in \cite{huang2014} and an example from \cite{lakomy2021tie}, where the reference signal $x_d(t)$ and fields $f^*(\cdot)$, $b^*(\cdot)$ are allowed to have a limited number of discontinuities with bounded left- and right-hand side limits.
\end{remark}

\begin{remark}
    In the specific context of ADRC, the synthesis of the control system expressed in the error domain has a certain beneficial characteristic; namely, consecutive time derivatives of $x_d(t)$ do not necessarily need to be known in advance. Details can be found in \cite{ErrorBasedADRCcep,CEPtorsionalFreq}.
\end{remark}

Now, a state vector $\pmb{e} = [e_1 \ e_2 \ \hdots \ e_n]^\top\triangleq[e \ \dot{e} \ \hdots \ e^{(n-1)}]^\top\in\mathbb{R}^n$ can be defined for the control error-domain system \eqref{eq:errorDomainSystem}, allowing its dynamics to be written in the following canonical observable form:
\begin{align}
    \begin{cases}
        \dot{\pmb{e}}(t) = \pmb{A}_n\pmb{e}(t) + \pmb{b}_n\left[f(\pmb{e},t) + b(\pmb{e},t)u(t) - d^*(t)\right], \\ 
        y(t) = \pmb{c}_n^\top \pmb{e}(t) - w(t),
    \end{cases}
    \label{eq:vectorizedErrorDomainSystem}
\end{align}
where $\pmb{x}_d\triangleq[x_d \ \dot{x}_d \ \hdots \ x^{(n-1)}_d]^\top\in\mathbb{R}^{n}$, $f(\pmb{e},t):=- f^*(x_d-e, \dot{x}_d - \dot{e}, \hdots, x_d^{(n-1)} - e^{(n-1)},t)$, and $b(\pmb{e},t):=- b^*(x_d-e, \dot{x}_d - \dot{e}, \hdots, x_d^{(n-1)} - e^{(n-1)},t)$. 
The general concept of robust control is usually connected with robustness against the structural and parametric uncertainties of the controlled system. The most conventional formulation of the minimal amount of information required for the application of ADRC can thus be formulated as the following assumption:
\begin{assumption}
    \label{ass:7}
    The structure and parameters of system \eqref{eq:vectorizedErrorDomainSystem} are highly uncertain up to a point where the dynamics order $n$ and a rough estimate $\hat{b}(\hat{\pmb{e}},t)\approx b(\pmb{e},t)$ are the only available information concerning the controlled system.
\end{assumption}

\begin{remark}
    \label{rem:3}
    The range of admissible ratio of $b(\cdot)/\hat{b}(\cdot)$, which needs to be satisfied to keep the ADRC-based system stable, was studied in detail in \cite{xue2018} and \cite{chen2020}, and is at least equal to 
    \begin{align}
        \forall_{t\geq0}\frac{b(\pmb{e},t)}{\hat{b}(\hat{\pmb{e}},t)}\in\left(0,2+\frac{2}{n}\right),\quad \textrm{for}~\hat{b}(\hat{\pmb{e}},t)\neq 0.
        \label{eq:gratio}
    \end{align}
\end{remark}

In the next step, following Assumption~\ref{ass:7}, system~\eqref{eq:vectorizedErrorDomainSystem} can be rewritten as
\begin{align}
    \begin{cases}
    \dot{\pmb{e}}(t) = \pmb{A}_n\pmb{e}(t) + \pmb{b}_n\left[\hat{b}(\hat{\pmb{e}},t)u(t) + d(\pmb{e},\hat{\pmb{e}},u,t)\right], \\ 
    y(t) = \pmb{c}_n^\top \pmb{e}(t) - w(t),
    \end{cases}
    \label{eq:knownDynamics}
\end{align}
where
\begin{align}
    d(\pmb{e},\hat{\pmb{e}},u,t) = -d^*(t) + f(\pmb{e},t) + \left[{b}({\pmb{e}},t) - \hat{b}(\hat{\pmb{e}},t)\right]u(t),
    \label{eq:totalDisturbance}
\end{align}
is the lumped total disturbance, which aggregates the unknown components that affect the system dynamics. 

To design the ADRC controller, an extended state $\pmb{z}\triangleq[\pmb{e}^\top \ d]^\top\in\mathbb{R}^{n+1}$ needs to be defined first, whose dynamics, based on \eqref{eq:knownDynamics}, can be written as
\begin{align}
    \begin{cases}
        \dot{\pmb{z}}(t) = \pmb{A}_{n+1}\pmb{z}(t) + \pmb{b}_{n+1}\dot{d}(\pmb{z},\hat{\pmb{z}},u,t) + \pmb{d}_{n+1}\hat{b}(\hat{\pmb{e}},t)u(t), \\ 
        y(t) = \pmb{c}^\top_{n+1}\pmb{z}(t) - w(t).
    \end{cases}
    \label{eq:extendedDynamics}
\end{align}
\begin{remark}
    Based on Assumptions \ref{ass:1}-\ref{ass:7}, it can be claimed that the derivative of the total disturbance is bounded, and $\dot{d}\in\mathcal{D}_{\dot{d}}\triangleq\{\dot{d}\in\mathbb{R}:|\dot{d}|<r_{\dot{d}}\}$ for some $r_{\dot{d}} > 0$.
\end{remark}

Because of equation~\eqref{eq:knownDynamics}, the estimates of vector $\hat{\pmb{z}} = [\hat{\pmb{e}}^\top \ \hat{d}]^\top$ (or its components) were used. These deliberate manipulations of introducing estimated signals into the system dynamics were indirectly caused by the form of system output, which allows measurement of only the first component of vector $\pmb{z}$ (see~\eqref{eq:extendedDynamics}), and thus, the rest of the extended state components have to be estimated using an extended state observer (ESO), which takes the form 
\begin{align}
    \dot{\hat{\pmb{z}}}(t) = \pmb{A}_{n+1}\hat{\pmb{z}}(t) + \pmb{d}_{n+1}\hat{b}(\hat{\pmb{z}},t)u(t) + \pmb{l}_{n+1}\left[y(t)-\pmb{c}^\top_{n+1}\hat{\pmb{z}}(t)\right],
    \label{eq:observer}
\end{align}
where the observer gain vector $\pmb{l}_{n+1} = [l_1 \ l_2 \ \hdots \ l_{n+1}]^\top\in\mathbb{R}^{n+1}$ is bandwidth-parametrised (see \cite{GaoPolePlacement}) with a single parameter $\omega_o>0$ such that 
\begin{align}
    l_i \triangleq \frac{(n+1)!}{i!(n+1-i)!}\omega_o^i, \quad \textrm{for} \  i\in\{1,...,n+1\}.
    \label{eq:parametrization}
\end{align}
Now, knowing how to calculate the values of $\hat{\pmb{z}}$, one can introduce a robust active disturbance rejection controller in the form of
\begin{align}
    u(t) \triangleq \frac{1}{\hat{b}(\hat{\pmb{e}},t)}\left[ - \pmb{k}_{n}\hat{\pmb{e}}(t) -  \hat{d}(t)\right],
    \label{eq:controller}
\end{align}
where $\pmb{k}_{n}=[k_1 \ k_2 \ \hdots \ k_{n}]\in\mathbb{R}^{1\times n}$. All controller gains are bandwidth-parametrised with a single parameter $\omega_c>0$ such that
\begin{align}
    k_i \triangleq  \frac{n!}{i!(n-i)!}\omega_c^i, \quad \textrm{for} \ i\in\{1,..,n\}.
    \label{eq:controllerParametrization}
\end{align}

\begin{remark}
    The presence of the control signal $u(t)$ in the total disturbance $d(\cdot)$, visible in \eqref{eq:totalDisturbance}, does not cause instability of the system as long as condition~\eqref{eq:gratio} is satisfied.
\end{remark}

After the application of~\eqref{eq:observer} and~\eqref{eq:controller} in \eqref{eq:extendedDynamics} and \eqref{eq:knownDynamics}, the subsystems that describe the dynamics of the observation error $\tilde{\pmb{z}}(t) \triangleq\pmb{z}(t) - \hat{\pmb{z}}(t)$ and the closed-loop control error $\pmb{e}(t)$ can be written as
\begin{align}
    \begin{cases}
        \dot{\tilde{\pmb{z}}}(t) = \left(\pmb{A}_{n+1}-\pmb{l}_{n+1}\pmb{c}_{n+1}\right)\tilde{\pmb{z}}(t) + \pmb{b}_{n+1}\dot{d}(t) - \pmb{l}_{n+1}w(t), \\ 
        \dot{\pmb{e}}(t) = \left(\pmb{A}_{n}-\pmb{b}_n\pmb{k}_n\right)\pmb{e}(t) + [\pmb{k}_{n} \ 1]\tilde{\pmb{z}}(t).
    \end{cases}
    \label{eq:errorDynamics}
\end{align}
The theoretical analysis of dynamical systems similar to \eqref{eq:errorDynamics} was studied in \cite{kicki2021tuning}, \cite{lakomy2021tie}, and \cite{lakomy2021kka}, and can be summarised with the following lemmas.

\begin{lemma}
    \label{lem:1}
    For $\omega_o>1$, and under Assumptions \ref{ass:1}-\ref{ass:7}, the observation error subsystem in~ \eqref{eq:errorDynamics}, parameterised by \eqref{eq:parametrization}, is input-to-state stable with respect to inputs $\dot{d}$ and $w$, and locally satisfies
    \begin{align}
        \|\tilde{\pmb{z}}(t)\|\leq\omega_o^nc_1\|\tilde{\pmb{z}}(0)\|\exp{\left(-c_2\omega_ot\right)}+\frac{1}{\omega_o}c_3\left(r_{\dot{d}}+\omega_o^{n+1}c_4r_w\right),
        \label{eq:lem1}
    \end{align}
    for some constants $c_1,c_2,c_3,c_4>0$.
\end{lemma}

\begin{lemma}
    \label{lem:2}
    For $\omega_c>1$, and under Assumptions \ref{ass:1}-\ref{ass:7}, the control error subsystem in \eqref{eq:errorDynamics}, parameterised by \eqref{eq:controllerParametrization}, is input-to-state stable and locally satisfies
    \begin{align}
        \|\pmb{e}(t)\| \leq \omega_c^{n-1}c_5\|\pmb{e}(0)\|\exp{\left(-c_6\omega_ct\right)} + \omega_c^{n-2}c_7\sup_{t\geq0}\|\tilde{\pmb{z}}(t)\|,
        \label{eq:lem2}
    \end{align}
    for some $c_5,c_6,c_7>0$.
\end{lemma}


\section{Functionalities and parameters of the ADRC Toolbox}
\label{sec:Overview}

\begin{figure}[t]
    \centering
    \includegraphics[width=0.5\textwidth]{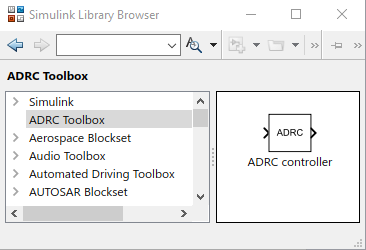}
    \caption{The proposed ADRC Toolbox (left) with the drag-and-drop ADRC function block (right), as seen in MATLAB\slash Simulink.}
    \label{fig:simpleDiagramMatlab}
\end{figure}

\begin{figure}[p]
    \centering
        \includegraphics[height=4.5cm]{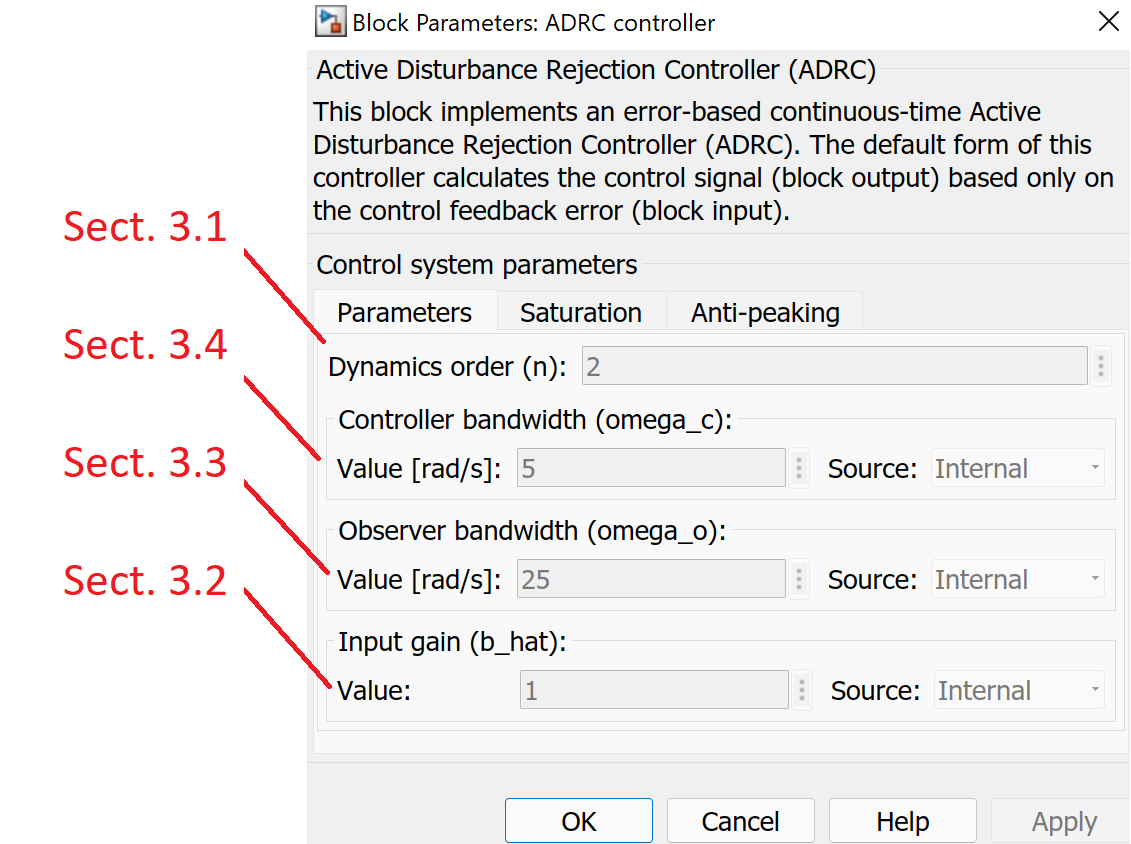}
        \includegraphics[height=4.5cm]{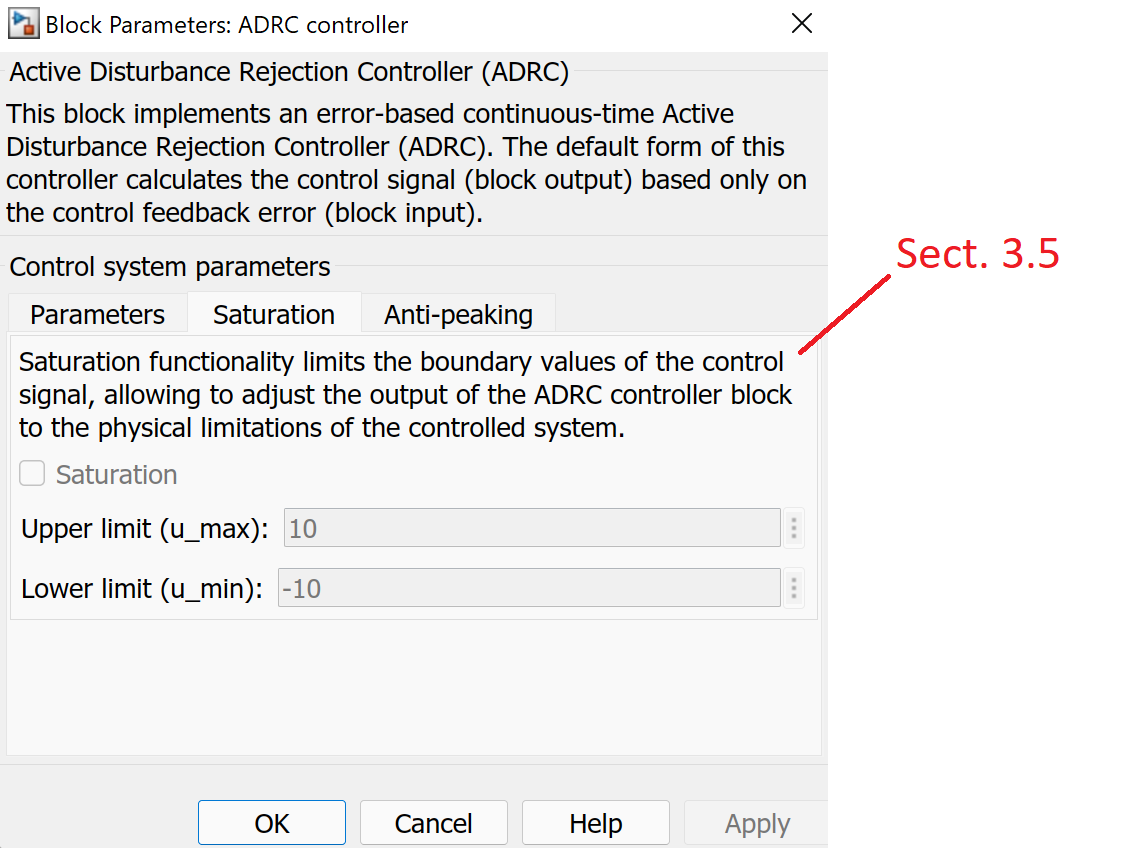}\\
        \includegraphics[height=4.5cm]{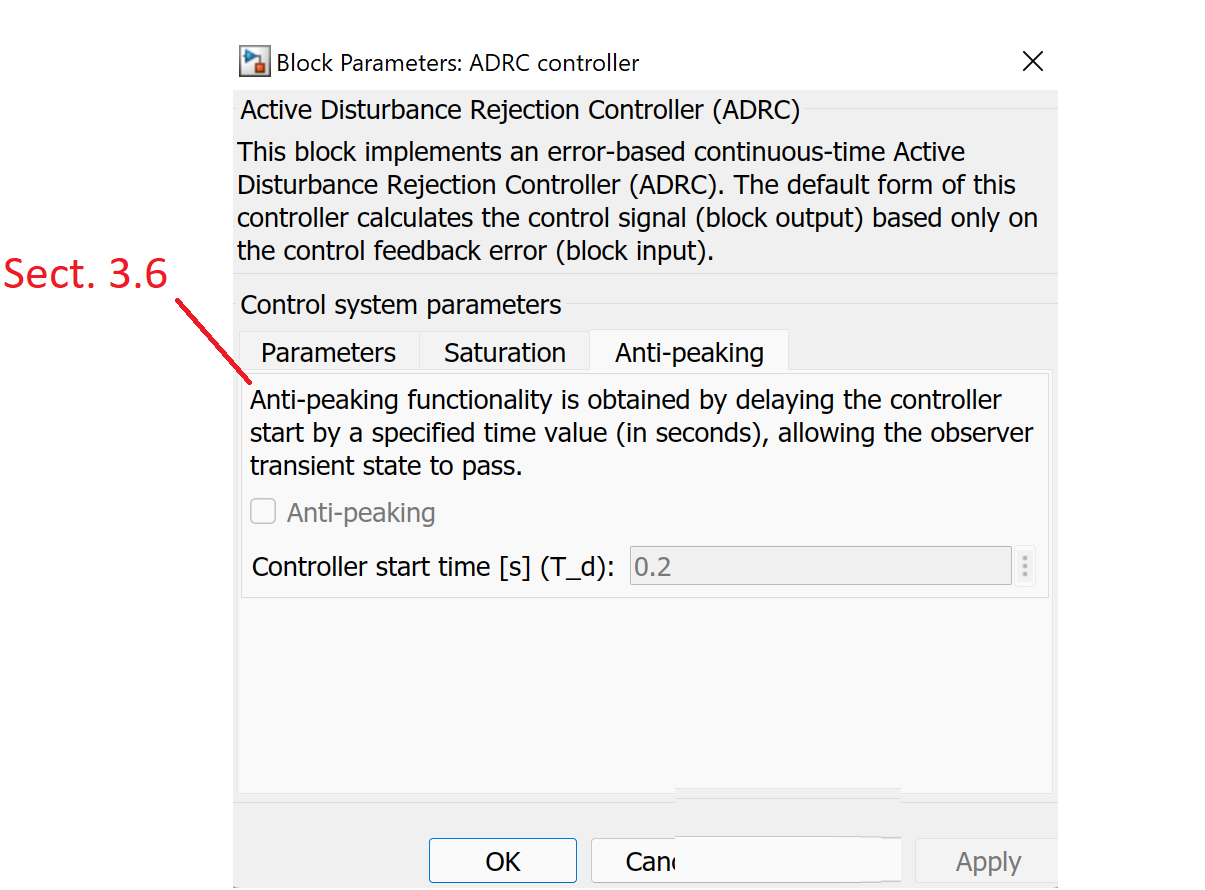}
    \caption{\textcolor{black}{The graphical user interface (GUI) for the proposed ADRC function block.}}
    \label{fig:GUI}
\end{figure}

The introduced ADRC Toolbox for MATLAB\slash Simulink contains a maximally simplistic, user-friendly, and general-purpose function block, that allows engineers to effectively use the ADRC concept, but with less knowledge of the method itself. The goal is to simplify the practical implementation of the architecture presented in Sect.~\ref{sec:ADRCpreliminaries}. The prepared drag-and-drop ADRC function block is a SISO block, where the input signal is the control error ($e$) and the output signal is the control signal ($u$). In the MATLAB\slash Simulink environment, the developed block is shown in Fig.~\ref{fig:simpleDiagramMatlab}, and it comes with the GUI presented in Fig.~\ref{fig:GUI}. 

In the described version of the ADRC function block (ver. 1.1.3), four mandatory parameters must be selected: 
\begin{itemize}
    \item the order of system dynamics $n$ -- see \eqref{eq:originalDynamics};
    \item an estimate of the input gain parameter $\hat{b}(\cdot)$ -- see \eqref{eq:knownDynamics};
    \item the observer bandwidth $\omega_o$ -- see \eqref{eq:observer} and \eqref{eq:parametrization}; and
    \item the controller bandwidth $\omega_c$ -- see \eqref{eq:controller} and \eqref{eq:controllerParametrization}.
\end{itemize}
Additionally, the ADRC function block enables two practically important functionalities, specifically, the control signal saturation and an anti-peaking mechanism. A thorough description of the aforementioned parameters and functionalities, together with the heuristics and interpretations concerning tuning, are provided in the following subsections.


\subsection{Order of system dynamics $n$}
\label{sec:Orderofsystem}

The dynamics order $n$ determines the dimensions of the observer state vector $\hat{\pmb{z}}$, and thus influences the number of observer equations. In practice, it is sometimes possible to achieve satisfactory control precision, even with an inappropriately set value $n$ for the ADRC controller (especially while following slowly varying trajectories)l; however, this is not recommended, because it can easily lead to system instability when changing other parameters. Hence, order $n$ should reflect the user's best knowledge of the relative order of the controlled system dynamics. 

From a theoretical point of view, according to Lemmas~\ref{lem:1} and~\ref{lem:2}, the higher the order $n$, the more fragile the control structure is with respect to measurement noise ($r_w$ multiplied by $\omega_o^{n+1}$), and the transient stage of the estimation error reaches larger values caused by observer peaking (first component of~\eqref{eq:lem1} multiplied by $\omega_o^n$). The order of the system dynamics also determines the range of acceptable input gain assessment $b(\cdot)/\hat{b}(\cdot)$, as  discussed in Remark~\ref{rem:3}. In the case of an unknown system order, a good starting point is setting $n$ to~1 or~2, because it represents many real physical systems with an acceptable level of approximation~\cite{bookADRC}.

\subsection{Estimate of input gain $\hat{b}(\cdot)$}
\label{sec:Inputgain}

The approximation range of $b(\cdot)$ is reasonably wide (see Remark~\ref{rem:3}), which implies a certain level of ADRC robustness with respect to the parametric uncertainty of $b(\cdot)$. However, as much precision as possible is recommended when setting the value $\hat{b}(\cdot)$. Not only can it lead to system instability when set outside the aforementioned acceptable range, but it can also cause the total disturbance to be more difficult to estimate, because the difference between the real input gain parameter $b(\cdot)$ and its assumed value $\hat{b}(\cdot)$ is incorporated within the total disturbance, represented by \eqref{eq:totalDisturbance}. Furthermore, when the value $\hat{b}(\cdot)$ approaches the boundaries of \eqref{eq:gratio}, unwanted oscillations may occur in the system behaviour. The input gain parameter $\hat{b}(\cdot)$ is also control signal scaling factor \eqref{eq:controller}. Thus, it also has an impact on the system performance when the admissible range of control signals is limited by a relatively narrow saturation set. 

\subsection{Observer bandwidth $\omega_o$}
\label{sec:Observerbandwidth}

With the chosen bandwidth-parameterisation described by~\eqref{eq:parametrization}, a single parameter, $\omega_o$, determines the values of all observer gains. Such an approach places each pole of the observation error dynamics state matrix (see~\eqref{eq:errorDynamics}) as
\begin{align}
    \textrm{eig}_i\left(\pmb{A}_{n+1} - \pmb{l}_{n+1}\pmb{c}_{n+1}^\top\right) = -\omega_o, \quad \textrm{for} \ i\in\{1,...,n+1\}.
    \label{eq:obsGain}
\end{align}
In addition to its simplistic form, this approach results in relatively good performance, even when compared to situations in which the eigenvalues are selected separately (see~\cite{kicki2021tuning}). 

According to the theoretical results from Lemma~\ref{lem:1}, relatively higher values of observer bandwidth have several effects on the system performance:
\begin{itemize}
    \item the estimation error of the extended state is less affected by the derivative of total disturbance influencing dynamics~\eqref{eq:extendedDynamics}, since the component connected with $\dot{d}$ is multiplied by $1/\omega_o$;
    \item the performance of the estimation system with respect to the measurement noise is highly influenced, because the impact of noise $w$ is multiplied by $\omega_o^n$; and
    \item the peaking phenomenon reaches higher values due to the multiplier of the first component in~\eqref{eq:lem1} equal to $\omega_o^n$, but does not last as long, because the speed of exponential decay that is connected with the peaking phenomenon is proportional to $\omega_o$. 
\end{itemize}

Here, it is worth mentioning that the observer bandwidth should be significantly larger than the controller bandwidth (i.e. $\omega_o\gg\omega_c$), because one of the controller goals is to compensate for the total disturbance $d(\cdot)$ which, if done properly, needs to be estimated in a much quicker manner. In \cite{GaoPolePlacement}, through simulation and experimentation, it was found that setting $\omega_o=5\raisebox{-0.9ex}{\~{}}10\omega_c$ is a reasonable compromise between the convergence speed and noise sensitivity in many control scenarios; hence, it is a good starting point for further fine-tuning. Moreover, as seen in Lemma~\ref{lem:2}, the precision of estimation (highly correlated with the value of $\omega_o$) determines the magnitude of the perturbation that affects the upper and lower bounds of the control error.

\subsection{Controller bandwidth $\omega_c$}
\label{sec:Controllerbandwidth}

In the process of controller tuning~\eqref{eq:controllerParametrization}, the same bandwidth-parameterisation concept is used as in the case of the observer, but this time, the eigenvalues of the state matrix are placed in the closed-loop control error subsystem \eqref{eq:errorDynamics}, that is,
\begin{align}
    \textrm{eig}_i\left(\pmb{A}_n-\pmb{b}_n\pmb{k}_n\right) = -\omega_c, \quad \textrm{for} \ i\in\{1,...,n\}.
    \label{eq:controlGain}
\end{align}
The increase in parameter $\omega_c$ causes a faster convergence of the control error from the initial conditions $\pmb{e}(0)$, but also possibly causes a higher overshoot because the first component of \eqref{eq:lem2} is multiplied by $\omega_c^{n-1}$. Looking at the second component of \eqref{eq:lem2}, higher values of $\omega_c$ may also have, depending on the dynamics order $n$, a negative impact on the influence of estimation errors on the control performance; thus, one should guarantee a fast and precise extended state estimation by selecting $\omega_o\gg\omega_c$ to reduce the impact of $\|\tilde{\pmb{z}}(t) \|$ on $\|\pmb{e}(t)\|$.

It is also worth noting that the transients of $d$ are faster with higher values of $\omega_c$, making it more difficult for the ESO to estimate them. Here again, a good starting point for controller bandwidth selection is the heuristic $\omega_o=5\raisebox{-0.9ex}{\~{}}10\omega_c$, as discussed in~\cite{GaoPolePlacement}.   

\subsection{Control signal saturation}
\label{sec:Outputsaturation}

When one knows the physical limitations of the controlled plant, enabling the control signal saturation function in the provided ADRC function block may have an enormous impact on the control performance. When this functionality is disabled (default setting), the observer treats unsaturated control $u$ as the signal that affects the system dynamics, whereas in reality, the dynamic system may be subject only to its saturated value. Such a situation may result in a virtual change in $b(\cdot)$ from the observer perspective and cause aggressive and unpredictable transients. In the worst case, it can lead to control system instability owing to the divergence from the range discussed in Remark~\ref{rem:3}. Depending on the selected saturation function status in the ADRC function block, the control signal takes the following form:
\begin{equation}
    u(t) = 
    \begin{cases}
    u^*(t), & \text{when saturation OFF (default)}; \\
    \begin{cases}
        u_{\text{min}}, & \textrm{for} \quad u^*(t) < u_{\text{min}}, \\ 
        u^*(t), &\textrm{for} \quad  u_{\text{min}} \leq u^*(t) \leq u_{\text{max}}, \\
        u_{\text{max}}, &\textrm{for} \quad u^*(t) > u_{\text{max}},
    \end{cases}
    & \text{when saturation ON},
    \end{cases}
    \label{eq:control_saturation}
\end{equation}
where the auxiliary variable $u^*(t)\triangleq\frac{1}{\hat{b}(\hat{\pmb{e}},t)}\left[ - \pmb{k}_{n}\hat{\pmb{e}}(t) -  \hat{d}(t)\right]$ is the originally defined control signal resulting from~\eqref{eq:controller}.

\begin{remark}
    \label{rem:saturation}
        It should be noted that introduction of the saturation function~\eqref{eq:control_saturation} in the ADRC algorithm, derived in its nominal form in Sect.~\ref{sec:ADRCpreliminaries}, is not consistent with Lemmas~\ref{lem:1} and~\ref{lem:2} (both of which consider saturation-less cases). Although true, the presence of a saturation function in disturbance observer-based control designs (especially in the context of stability) has already been addressed in the literature (e.g., \cite{PerfRecovery,ADRCinputSat}). Additionally, the saturation of the control signal is most frequently encountered in the initial transient stage of the control experiments, after which the control signal does not reach saturation and acts according to~\eqref{eq:controller}, which is in line with Lemmas~\ref{lem:1} and~\ref{lem:2}.
\end{remark}

\subsection{Anti-peaking}
\label{sec:Antipeaking}

\textcolor{black}{The existence of the peaking phenomenon in the operation of high-gain observers is a well-known issue that can reduce the control system performance, especially at the beginning of the observation process~\cite{KhalilPralyHigGain}}. Although many methods can be used within the observer structure to address this issue (e.g. \cite{LowPower,SatCascade}), here, a method that deals with the peaking by temporarily disabling the control action at the beginning of the control process (i.e. the control signal is being calculated but not applied to the controlled plant), allows the peaking to simply pass. This approach was successfully used in \cite{lakomy2021isat,lakomy2021tie}. The specific anti-peaking logic implemented in the ADRC function block affects the control signal in the following manner:
\begin{equation}
    u(t) = 
    \begin{cases}
    u^*(t), & \text{when anti-peaking OFF (default)}; \\
    \begin{cases}
    0~\text{for}~0 \leq t \leq T_d; \\
    u^*(t)~\text{for}~t>T_d; 
    \end{cases} & \text{when anti-peaking ON},
    \end{cases}
    \label{eq:antipeakfcn}
\end{equation}
where $T_d>0$ is the time that must pass before applying the control signal, and $u^*(t)$ is the auxiliary variable that represents the control signal after passing the saturation logic defined in~\eqref{eq:control_saturation}.

\begin{remark}
    \label{rem:antipeaking}
        Here, similar to Remark~\eqref{rem:saturation}, which discusses the control signal saturation, the use of the anti-peaking mechanism~\eqref{eq:antipeakfcn} in the considered ADRC algorithm is inconsistent with Lemmas~\ref{lem:1} and~\ref{lem:2} (both derived for the case without anti-peaking). The theoretical ramifications of this have been studied before, for example, in~\cite{SatCascade,LowPower}, where it was shown that under certain conditions, which are relatively easy to satisfy in practice, the stability of the system can be retained.
\end{remark}

\subsection{Overview}

\begin{figure}
    \centering
    \includegraphics[width=0.55\textwidth]{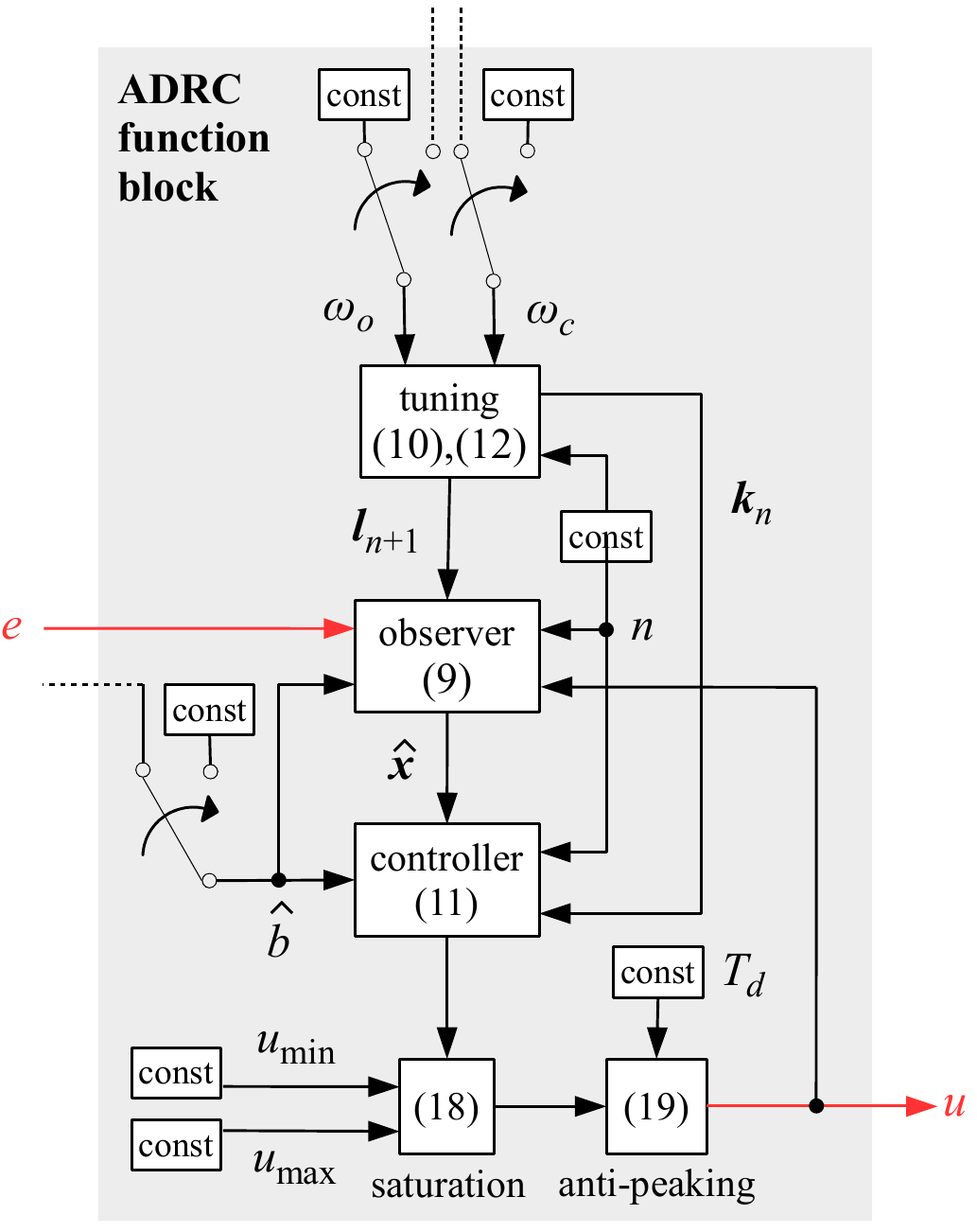}
    \caption{Schematic diagram of the ADRC function block (available in the developed ADRC Toolbox; cf.~Fig.~\ref{fig:simpleDiagramMatlab}).}
    \label{fig:example}
\end{figure}

A simplified diagram showing the insights of the introduced ADRC function block is shown in Fig.~\ref{fig:example}, and the scope of its current functionalities is summarised in Table~\ref{tab:ToolboxFunctions}. Depending on the function block configuration, it can be seen that $\hat{b}$ can be chosen to be either constant (realised internally through the provided GUI, Fig.~\ref{fig:GUI}) or supplied externally as a time-dependent value from outside the function block, which is illustrated by the switching mechanisms in Fig.~\ref{fig:example}. The default configuration in the ADRC function block is the constant $\hat{b}$. One can change its source in the GUI to external and then provide (in each sample time) information about the varying value of $\hat{b}$. A variety of techniques can be used to calculate $\hat{b}(t)$ on-line, for example, a genetic algorithm~\cite{PiosikMED} or extremum-seeking algorithm~\cite{ExtremumSeeking}. Regardless of the chosen type of $\hat{b}$ in the function block, its value must satisfy~\eqref{eq:gratio}. In the cases of $\omega_c$ and $\omega_o$, a similar approach is applied. By default, these values are to be set as a constant in the ADRC function block, or alternatively provided as external time-dependent values. In the latter case, their specific values can be calculated, for example, using a neural network~\cite{kicki2021tuning}, genetic algorithm~\cite{MultiobjOptimizer}, nonlinear function~\cite{HanTIE}, or time-varying~\cite{WeiTADRC} function.

\begin{figure}[p]
    \centering
    \includegraphics[height=0.95\textheight]{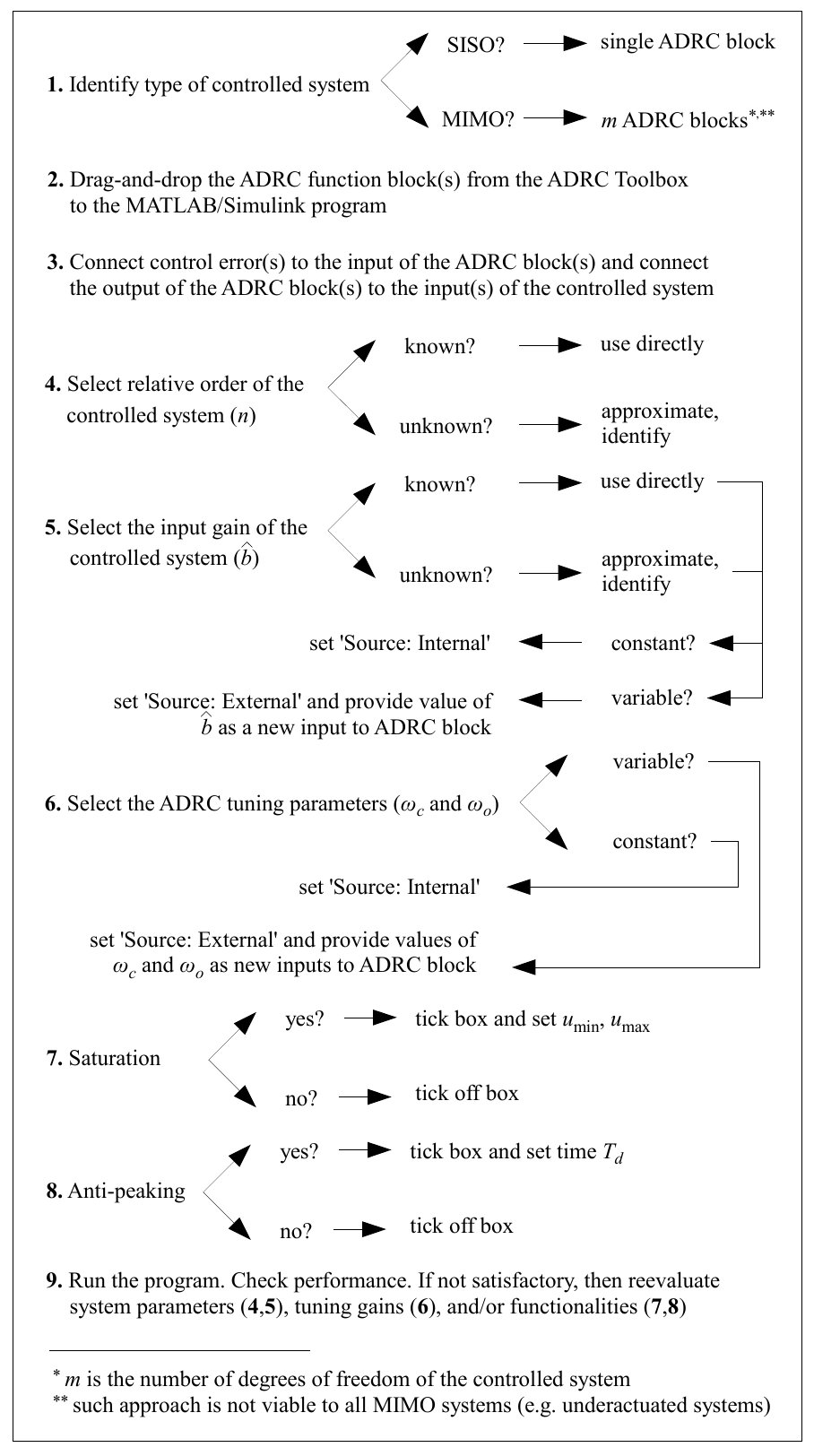}
    \caption{\textcolor{black}{Design steps for deploying the proposed ADRC function block.}}
    \label{fig:ADRCtoolboxDecisionScheme}
\end{figure}

To summarise, Fig.~\ref{fig:ADRCtoolboxDecisionScheme} shows the design steps for deploying the proposed ADRC function block. 

\begin{table}[]
\centering
\caption{Overview of parameters and functionalities of the ADRC function block.}
\label{tab:ToolboxFunctions}
\resizebox{\textwidth}{!}{%
\begin{tabular}{|c|c|c|c|c|}
\hline
\textbf{Parameter/Functionality}        & \textbf{Symbol}             &    \textbf{Selection}       & \textbf{Variations}                                & \textbf{Admissible value}     \\ \hline
Order of system dynamics (Sect.~\ref{sec:Orderofsystem})       & $n$                & Mandatory & Constant                                          & $n\in\mathbb{Z}$       \\ \hline
Estimate of input gain (Sect.~\ref{sec:Inputgain})           & $\hat{b}$          & Mandatory & Constant / Time-dependent & $\hat{b}\in\mathbb{R}/\{0\}$ \\ \hline
Observer bandwidth (Sect.~\ref{sec:Observerbandwidth})   & $\omega_o$         & Mandatory & Constant / Time-dependent & $\omega_o\gg\omega_c$ \\ \hline
Controller bandwidth (Sect.~\ref{sec:Controllerbandwidth}) & $\omega_c$         & Mandatory & Constant / Time-dependent & $\omega_c>0$ \\ \hline
Control signal saturation (Sect.~\ref{sec:Outputsaturation})    & n/a   & Optional  & Upper limit, Lower limit                   & $u_{\text{min}}<u_{\text{max}}$ \\ \hline
Anti-peaking (Sect.~\ref{sec:Antipeaking})         & n/a & Optional  & Constant                                   & $T_d>0$  \\ \hline
\end{tabular}%
}
\end{table}

\section{Validation}
\label{sec:Validation}

To verify the efficacy of the proposed ADRC Toolbox with its ADRC function block, a set of tests was conducted in Examples~\#1 through~\#5. The tests, described in Table~\ref{tab:ValidationTests}, utilise a group of systems from various control areas that differ from each other in terms of order, type, and structure of the controlled dynamics as well as type of external disturbance. Some of the control examples were performed in simulation, whereas others were realized in hardware. In contrast to simulation studies, physical benchmarks also consider real process characteristics such as discrete sampling intervals, communication overhead related to the process, requirements to meet a cycle time, and mathematical modelling mismatches. Such diversified group was purposefully selected to test the ADRC Toolbox in various control scenarios. Moreover, the experimental part of the validation was conducted using only hardware components that are relatively cheap and easily accessible. The choice of testing the proposed ADRC Toolbox exclusively using commercial off-the-shelf hardware systems was deliberate because it allows the Toolbox users to reproduce the results shown later in this paper. The tuning methodology in all the upcoming examples is based on the guidelines from Sect.~\ref{sec:Overview}.

\begin{table}[]
\centering
\caption{Methodology of ADRC Toolbox validation.}
\label{tab:ValidationTests}
\resizebox{\textwidth}{!}{%
\begin{tabular}{|c|c|c|c|c|c|c|c|}
\hline
\multirow{2}{*}{\textbf{Example no.}} &
\multirow{2}{*}{\textbf{Plant}} & \multicolumn{6}{c|}{\textbf{Criterion}} \\ \cline{3-8} 
 &  & \textbf{Order} ($n$) & \textbf{Type} & \textbf{Structure} &  \textbf{Ext. disturb.} ($d^*$) & \textbf{Control area} & \textbf{Validation} \\ \hline
\#1 (Sect.~\ref{sec:Example1}) & Generic plant & 4 & Linear & SISO & Step & n/a & Simulation \\ \hline
\#2 (Sect.~\ref{sec:Example2}) & Coupled tanks & 1 & Nonlinear & MIMO & Step & Process & Simulation \\ \hline
\#3 (Sect.~\ref{sec:Example3}) & Power converter & 2 & Linear & SISO & Harmonic & Power & Simulation \\ \hline
\#4 (Sect.~\ref{sec:Example4}) & DC motor & 2 & Linear & SISO & None & Motion & Experiment \\ \hline
\#5 (Sect.~\ref{sec:Example5}) & Heaters & 1 & Nonlinear & MIMO & None & Process & Experiment \\ \hline
\end{tabular}%
}
\end{table}

\subsection{Example \#1: Generic plant control}
\label{sec:Example1}

%
%
%

The first considered system is expressed in form of~\eqref{eq:originalDynamics} as 
\begin{align}
    \begin{cases}
        x^{(4)}(t) &= f^*\left(x, \dot{x}, \ddot{x}, x^{(3)}, t\right) + b^*u(t) + d^*(t), \\ 
        y^*(t) &= x(t) + w(t),
    \end{cases}
    \label{eq:example0dynamics}
\end{align}
where $f^*\left(x, \dot{x}, \ddot{x}, x^{(3)}, t\right) = -4x^{(3)} - 6\ddot{x} - 4\dot{x} - x$, $b^* = 1$, and $d^*(t) = 70\cdot \mathbbm{1}(t-10)$.

The control objective here is to generate the control signal $u$ so the output signal $x$ tracks the desired trajectory $x_d$, despite the presence of an external disturbance and the parametric uncertainty of the model. Additionally, two scenarios are considered in this study: without sensor noise ($w(t)\equiv0$) and with sensor noise $w(t)\sim \mathcal{N}\left(0, 10^{-12}\right)$. The ADRC function block (from the proposed ADRC Toolbox) is implemented with design parameters $n=4$, $\hat{b}=0.8$, $\omega_c=5$, $\omega_o=50$, and the anti-peaking functionality turned off (default setting). Two additional scenarios are considered: with the saturation turned off (default setting) and with the saturation turned on, with lower and upper limits $u_{\text{min}}=-100$ and $u_{\text{max}}=100$. With~\eqref{eq:example0dynamics} expressed in the form of~\eqref{eq:originalDynamics} and the ADRC parameters and functionalities configured, the remaining equations of the ADRC algorithm \eqref{eq:errorDomainSystem}-\eqref{eq:errorDynamics} can now be straightforwardly derived for the above system, which is omitted here to avoid redundancy.

\begin{figure}[t]
    \centering
    \includegraphics[width=0.75\textwidth]{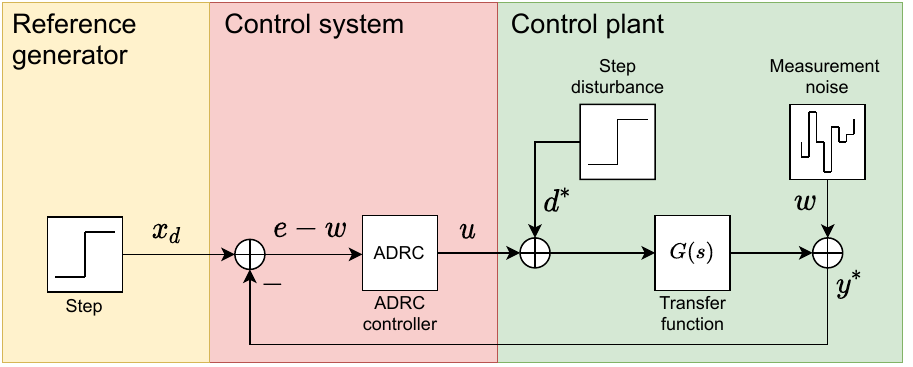}
    \caption{Block diagram of the generic control system (Example \#1) with the ADRC function block from the proposed ADRC Toolbox.}
    \label{fig:example1}
\end{figure}

\begin{figure}[t]
 \centering
    \begin{subfigure}{\textwidth}
        \includegraphics[width=0.32\textwidth]{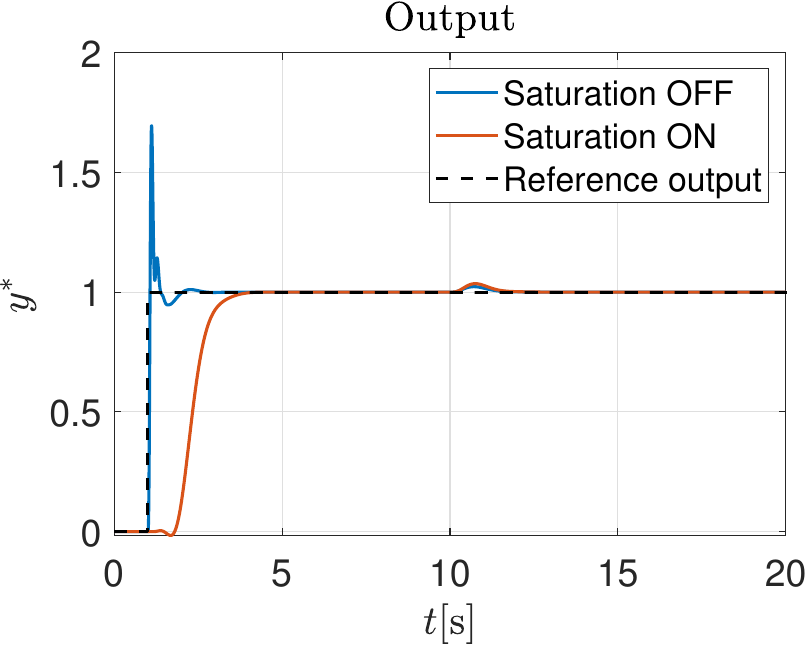}
        \includegraphics[width=0.32\textwidth]{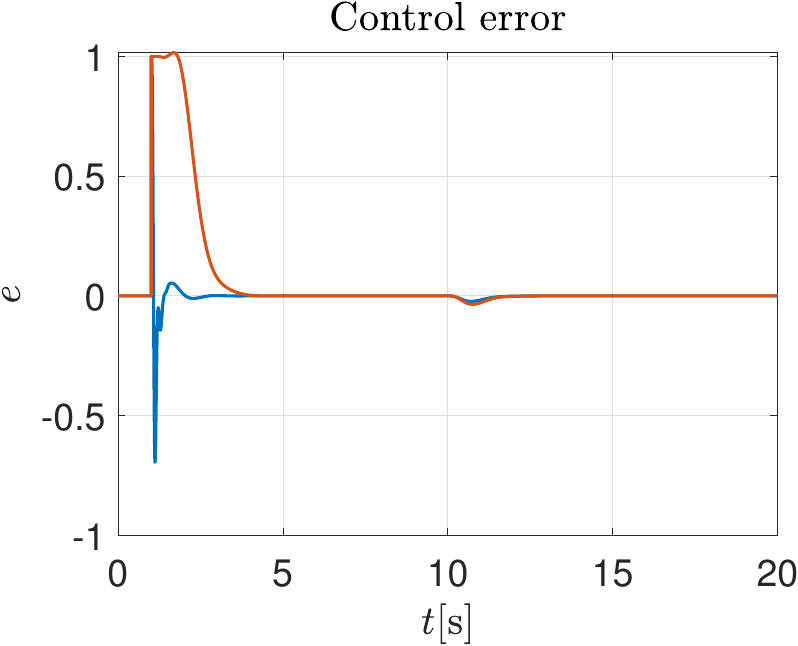}
        \includegraphics[width=0.32\textwidth]{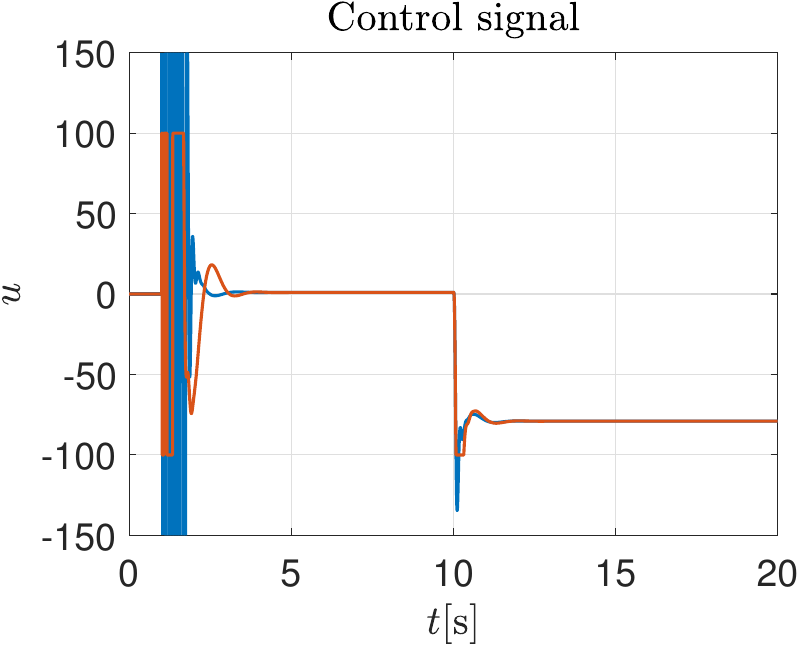}
        \caption{case without sensor noise ($w(t)\equiv0$)}
        \label{fig:results_e1_no_noise}
    \end{subfigure}
    \hfill
    \begin{subfigure}{\textwidth}
        \includegraphics[width=0.32\textwidth]{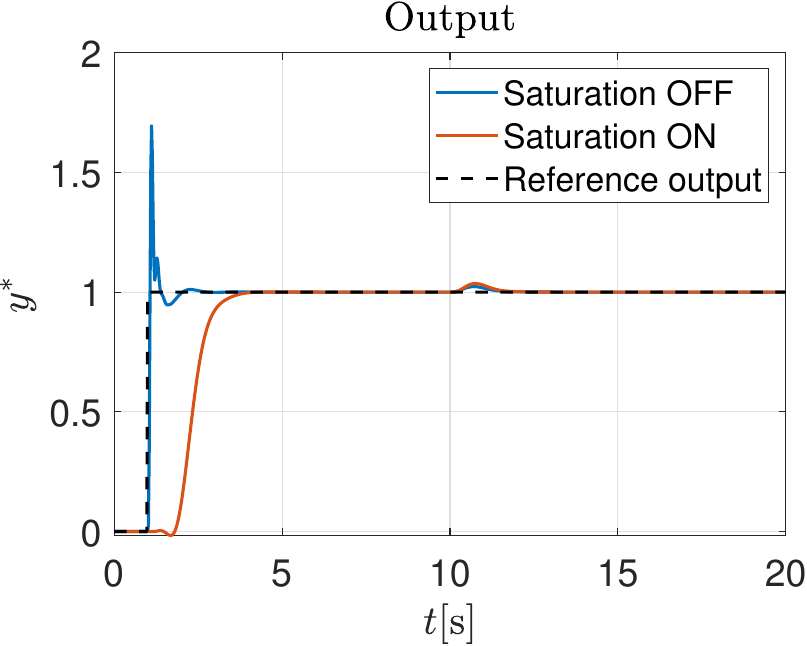}
        \includegraphics[width=0.32\textwidth]{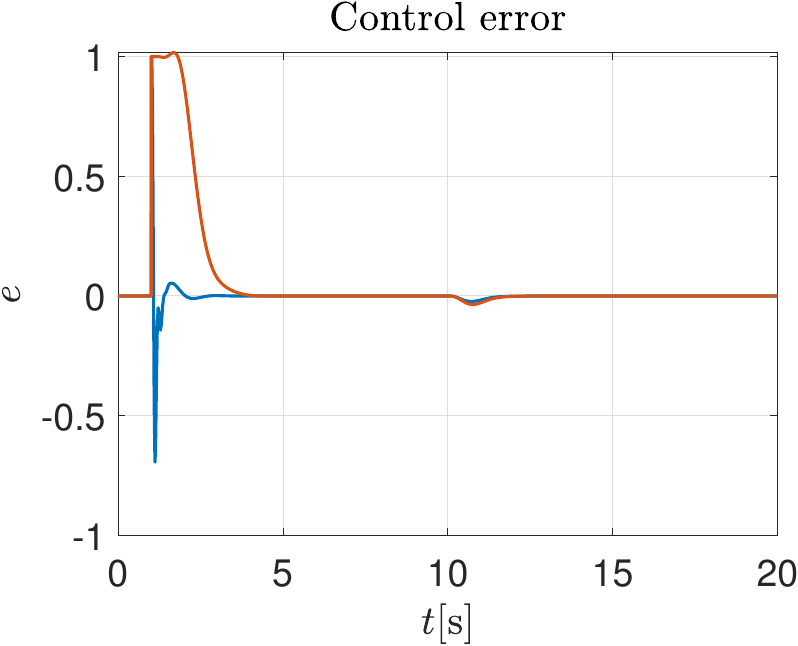}
        \includegraphics[width=0.32\textwidth]{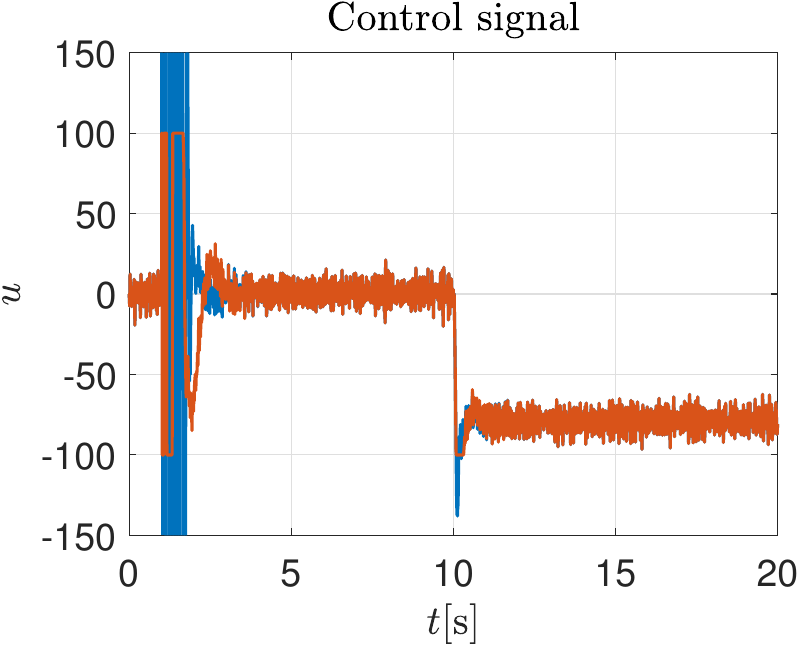}
        \caption{case with sensor noise $w(t)\sim \mathcal{N}\left(0, 10^{-12}\right)$}
        \label{fig:results_e1_noise}
    \end{subfigure}
        \caption{Simulation results of Example \#1.}
        \label{fig:results_e1}
\end{figure}

The block diagram of the implemented ADRC-based system in MATLAB\slash Simulink is shown in Fig.~\ref{fig:example1}, where $G(s)=1/\left(s^4+4s^3+6s^2+4s+1\right)$. The results of using the proposed ADRC Toolbox in the above control problem are shown in Fig.~\ref{fig:results_e1}. From the obtained results, one can conclude that the applied ADRC function block managed to realise the given control objective in a satisfactory manner for all tested scenarios. The implemented robust controller drove the control error toward zero and performance recovered after the appearance of the external disturbance at $t=10$s. One can also see the influence of the saturation mechanism and the measurement noise. In the case of the saturation, the control error converges faster when the saturation is off, except for the effects of the initial oscillatory behaviour of the output and control signal peaks, reaching the magnitude of $u\approx 10^6$ in the initial transient, which would probably not be feasible for most real-world controllers. In the case of measurement noise, owing to the particular structure of the observer and the controller (both utilising the output signal), the noise is clearly manifested in the control signal. At the same time, for the considered system, the noise in the control signal is mostly filtered by the plant dynamics and does not have a significant impact on the controlled variable.

\subsection{Example \#2: Coupled tanks level control}
\label{sec:Example2}


This example considers a hydraulic system, which consists of two tanks connected by a flow channel and two independent inlet flows, one for each tank. The fluid levels in the tanks are considered as the system outputs; hence, the resultant multi-input, multi-output (MIMO) system model can be expressed as
\begin{align}
    \begin{cases}
    \frac{dh_1(t)}{dt} &= -\frac{a}{c}\sqrt{2gh_1(t)} - \frac{a}{c}\text{sign}(\epsilon_h(t))\sqrt{2g|\epsilon_h(t)|} + \frac{1}{c}u_1(t) + d^*_1(t), \\
    y_1^*(t) &= h_1(t) + w_1(t), \\
    \frac{dh_2(t)}{dt} &= -\frac{a}{c}\sqrt{2gh_2(t)} + \frac{a}{c}\text{sign}(\epsilon_h(t))\sqrt{2g|\epsilon_h(t)|} + \frac{1}{c}u_2(t) + d_2^*(t), \\
    y_2^*(t) &= h_2(t) + w_2(t), 
    \end{cases}
    \label{eq:tanksModelMIMO}
\end{align}
where $u_1$[m$^3$/s] and $u_2$[m$^3$/s] are the inlet flows (control signals) for the first and second tanks, respectively; \textcolor{black}{$y_1^*$[m] and $y_2^*$[m] are the measured system outputs, which consist of, respectively, first and second tank fluid levels $h_{1}$[m] and $h_{2}$[m]} and the corresponding sensor noises $w_1$[m] and $w_2$[m]; the auxiliary variable $\epsilon_h:=h_1-h_2$ represents the fluid level difference; $g$[m/s$^2$] is the gravitational acceleration; $a$[m$^2$] is the cross-sectional area of the pipes; and $c$[m$^2$] is the cross-sectional area of the tanks. It is assumed that the system model~\eqref{eq:tanksModelMIMO} has the same parameters for both tanks and pipelines. The parameters of the system model used in the simulation were  $c=1.2\cdot 10^{-2}$m$^2$, $a=7.5\cdot 10^{-5}$m$^2$, and $g = 9.81$m/s$^2$. Referring to the generalised form of the system description from~\eqref{eq:originalDynamics}, one can assign $x(t):=h_1(t)$, $f^*(x,t) = -\frac{a}{c}\sqrt{2gx(t)} - \frac{a}{c}\text{sign}(\epsilon_h(t))\sqrt{2g|\epsilon_h(t)|}$, and $b^* = \frac{1}{c}$ for the first tank dynamics and $x(t):=h_2(t)$, $f^*(x,t) = -\frac{a}{c}\sqrt{2gx(t)} + \frac{a}{c}\text{sign}(\epsilon_h(t))\sqrt{2g|\epsilon_h(t)|}$, and $b^* = \frac{1}{c}$ for the second tank dynamics.

The control objective here is to manipulate two inlet flows $u_{1/2}$ to make the output levels $h_{1/2}$ track the respective reference set points $x_{d1/d2}$[m]. The control process should be carried out despite external disturbances $d_1^*(t) = -1.4\cdot 10^{-4}\cdot \mathbbm{1}(t-90)$ and $d_2^*(t) = -0.7\cdot 10^{-4}\cdot \mathbbm{1}(t-130)$, the parametric uncertainty of the model, and Gaussian sensor noise $w_{1/2}\sim \mathcal{N}\left(0, 0.8\cdot 10^{-10}\right)$. It is worth noting that, in this example, the tanks are treated as two independent systems; therefore, two independent ADRC function blocks are used, one to govern each input-to-output channel of the MIMO system. In such a configuration, the influence of the cross-couplings is treated as part of the total disturbance of the system.

\begin{figure}[t]
    \centering
    \includegraphics[width=0.8\textwidth]{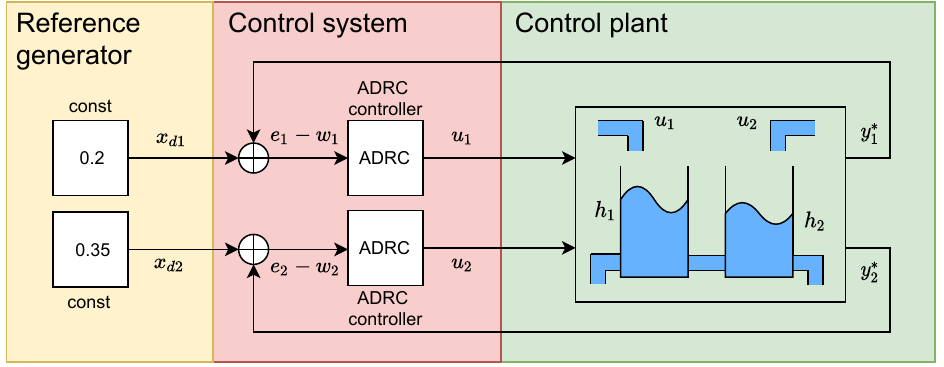}
    \caption{Block diagram of the coupled tank control system (Example \#2) with the ADRC function blocks (from the proposed ADRC Toolbox), each controls one degree of freedom.}
    \label{fig:example2}
\end{figure}

\begin{figure}[t]
    \centering
        \includegraphics[width=0.32\textwidth]{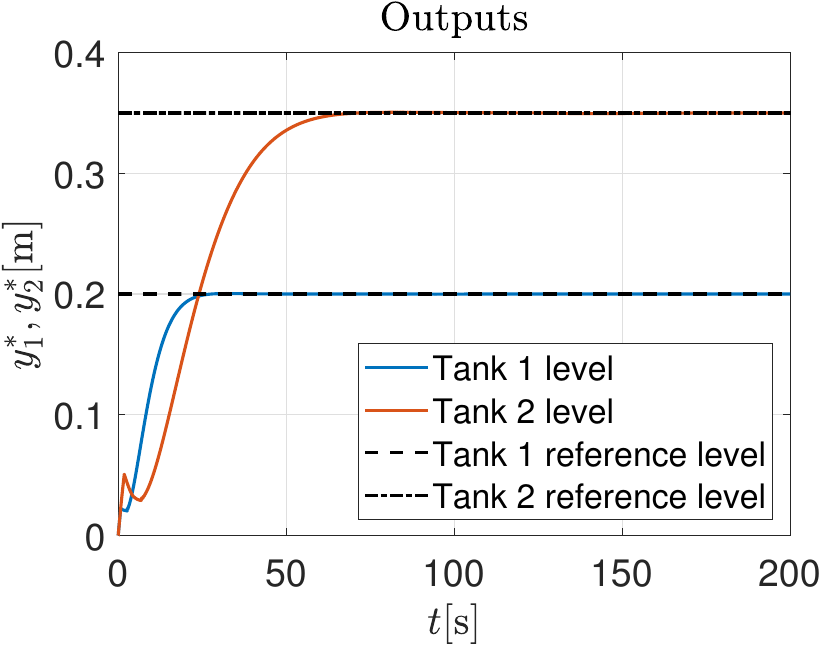}
        \includegraphics[width=0.32\textwidth]{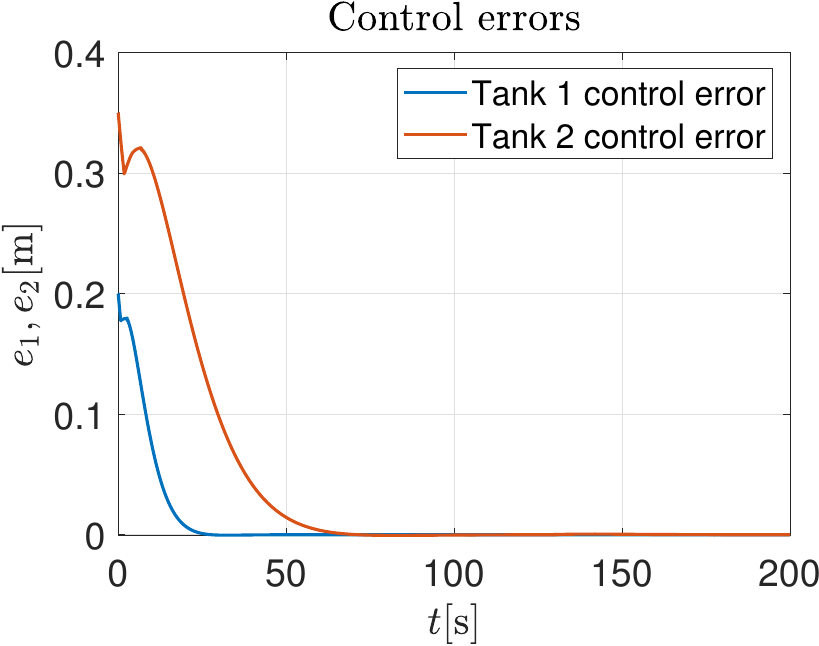}
        \includegraphics[width=0.32\textwidth]{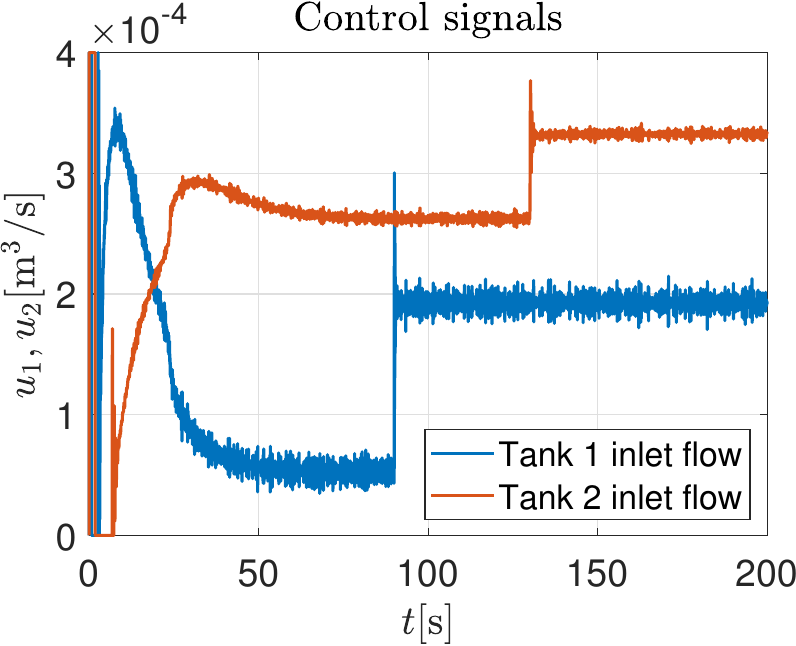}
    \caption{Simulation results of Example \#2.}
    \label{fig:results_tanks}
\end{figure}

In this example, the two ADRC blocks are implemented with design parameters $n=2$, $\hat{b}=0.8$, $\omega_c=0.3$, and $\omega_o=3$ for the first tank, and $n=2$, $\hat{b}=1$, $\omega_c=0.12$, and $\omega_o=1.2$ for the second tank. The control signal saturation is turned on with limits $u_{\text{min}}=0$ and $u_{\text{max}}=4\cdot 10^{-4}$ \textcolor{black}{(where the latter value corresponds to the fluid flow rate of $1440$l/h)} and the anti-peaking function is turned off (default setting). The block diagram of the implemented ADRC-based system in MATLAB\slash Simulink is shown in Fig.~\ref{fig:example2}, and the results of using the ADRC Toolbox in the above control problem are shown in Fig.~\ref{fig:results_tanks}. 

In this figure, one can see that the ADRC scheme successfully realises the control task in both control channels of the considered multidimensional system. After the transient stages, both tank levels are maintained at the desired predefined values, and the implemented ADRC controllers manage to quickly and accurately compensate for the influence of external disturbances before they have a visible effect on the process variables. However, the impact of $d_1^*$ and $d_2^*$ is clearly seen in the control signals $u_{1/2}$ at $t=90$s and $t=130$s, when they are abruptly forced by the control structure to generate more energy to handle the disturbances in real time. Interestingly, the ADRC block parameters for governing the first tank dynamics have higher tuning parameter values $\omega_c$ and $\omega_o$ and lower estimated input gain values $\hat{b}$, when compared to those for the second tank. This results in faster convergence of the control error for the first tank, but also creates larger noise amplification in the control signal, which could be potentially problematic in real-world implementations. These observations directly correspond to the theoretical findings described for $\hat{b}$, $\omega_o$, and $\omega_c$ in Sects.~\ref{sec:Inputgain}, \ref{sec:Observerbandwidth}, and \ref{sec:Controllerbandwidth}, respectively.

\subsection{Example \#3: Power converter voltage control}
\label{sec:Example3}

The system is a DC-DC buck converter, which is a common component in power electronics. Following~\cite{lakomy2021tie,CEPconverter}, the \textit{average} second-order dynamic model can be formulated as
	\begin{equation}
		\begin{cases}
		\frac{d^2v_o(t)}{dt^2} &= -\frac{1}{CR}\frac{dv_o(t)}{dt} - \frac{1}{CL}v_o(t)+\frac{V_{\text{in}}}{CL}u(t)+d^*(R_L(t),t),\\
		y^*(t) &= v_o(t) + w(t),
		\label{eq:DCDCmodel}
			\end{cases}
	\end{equation}
where $u \in [0,1]$ is the duty ratio (control signal); $y^*$[V] is the measured system output, which consists of the average capacitor voltage $v_o$[V] and the sensor noise $w$[V]; $R$[$\Omega$] is the load resistance of the circuit; $L$[H] is the filter inductance; $C$[F] is the filter capacitance; $V_{\text{in}}$[V] is the input voltage source; and the external disturbance $d^*$ is dependent on the time-varying load resistance $R_L$[$\Omega$]. The parameters of the power converter used in the simulation were taken from a real system that was utilised in~\cite{lakomy2021tie, CEPconverter} and are $V_{\text{in}}=20$V, $L=0.01$H, $C=0.001$F, and $R=50\Omega$. Referring to the generalised form of the system description in \eqref{eq:originalDynamics}, we can assign $x(t):=v_o(t)$, $f^*(x,\dot{x},t) = -\frac{1}{CR}\dot{x}(t) - \frac{1}{CL}x(t)$, and $b^* = \frac{V_{\text{in}}}{CL}$.

The control objective here is to generate the control signal $u$ that will make $v_o$ track a reference capacitor output voltage trajectory $x_d$[V] despite the unknown external disturbance related to load resistance $R_L = 85\sin(40\pi t) + 100$, the presence of sensor noise $w\sim\mathcal{N}(0,5\cdot10^{-5})$, and the parametric uncertainty of the internal model dynamics.

\begin{figure}
    \centering
    \includegraphics[width=1.0\textwidth]{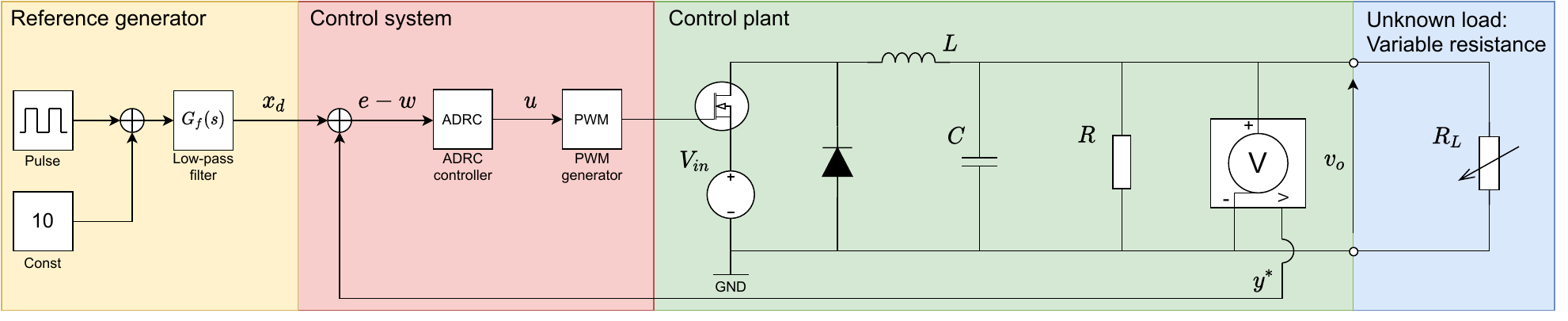}
    \caption{Block diagram of the DC-DC buck power converter control system (Example \#3) with the ADRC function block from the proposed ADRC Toolbox.}
    \label{fig:example3}
\end{figure}

\begin{figure}
    \centering
    \includegraphics[width=0.32\textwidth]{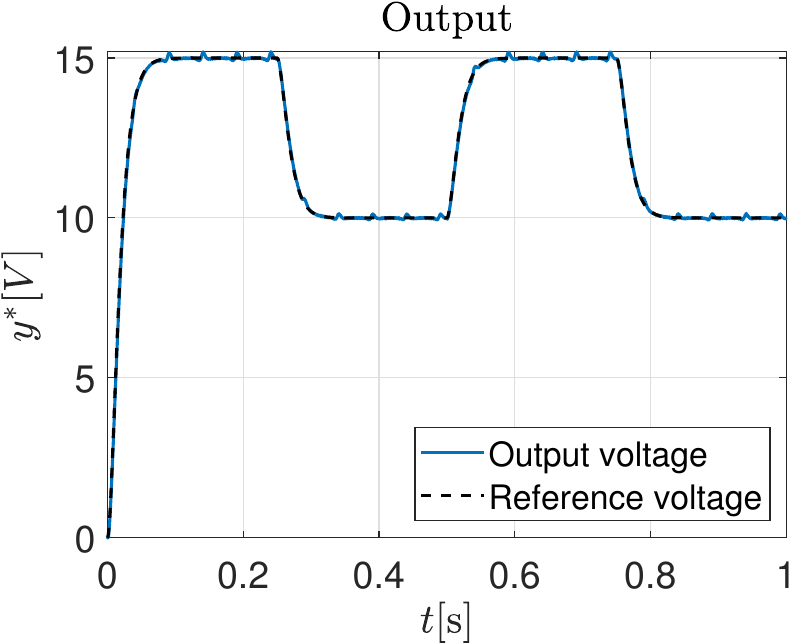}
    \includegraphics[width=0.32\textwidth]{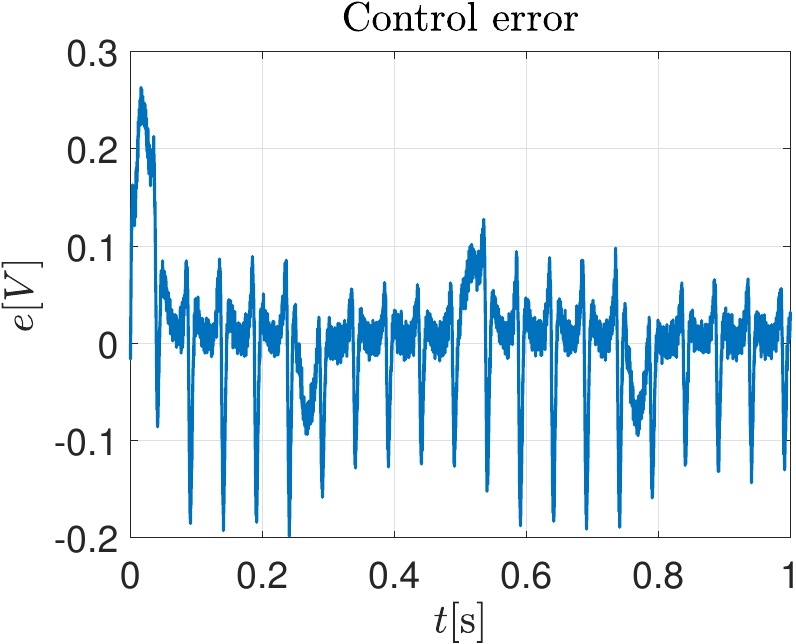}
    \includegraphics[width=0.32\textwidth]{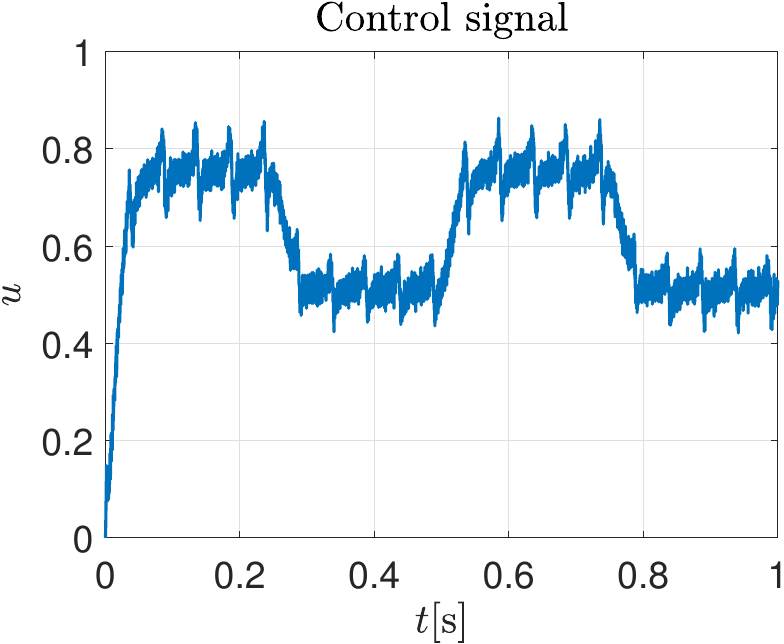}\\
    \includegraphics[width=0.32\textwidth]{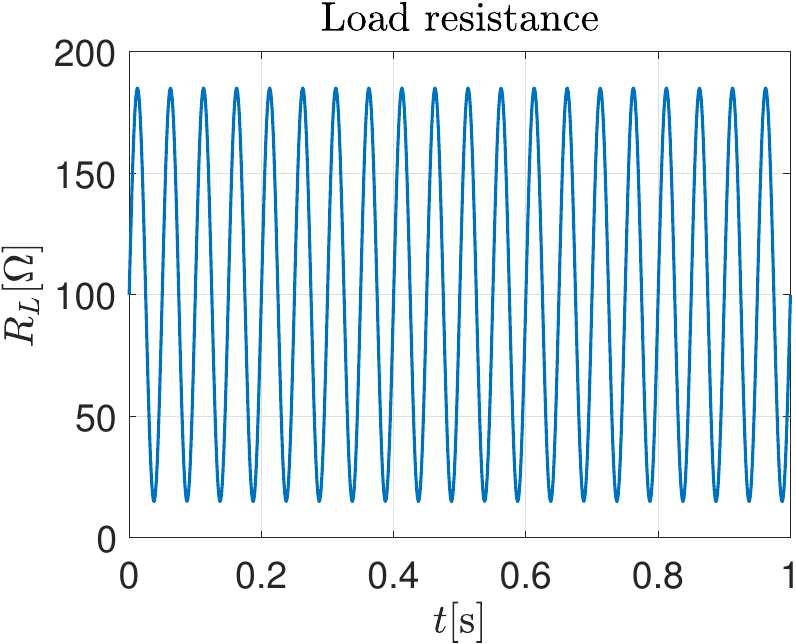}
    \includegraphics[width=0.32\textwidth]{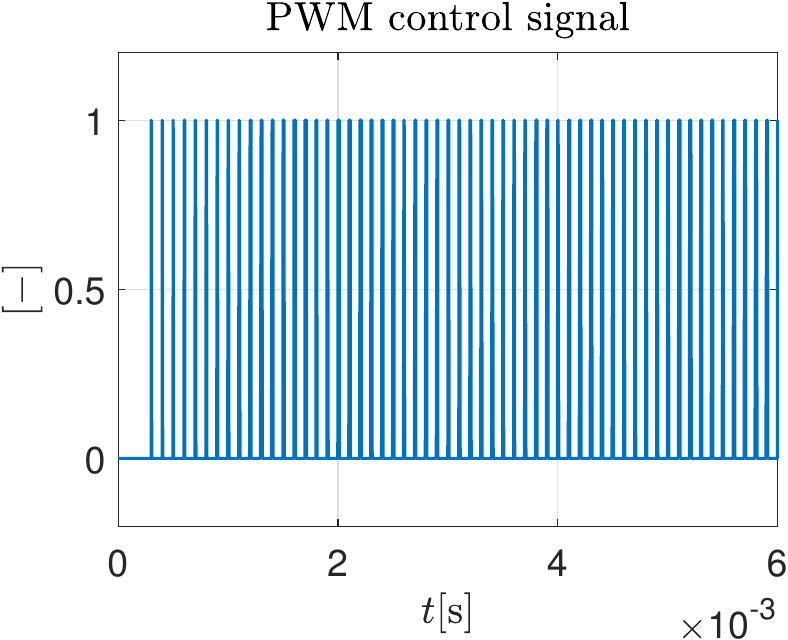}
    \caption{Simulation results of Example \#3.}
    \label{fig:results_converter}
\end{figure}

In this example, the ADRC function block is implemented with the design parameters $n=2$, $\hat{b}=2\cdot10^{6}$, $\omega_c=500$, and $\omega_o=3000$, the anti-peaking mechanism turned off (default setting), and the saturation of the control signal turned on, with $u_{\text{min}}=0$ and $u_{\text{max}}=1$. A block diagram of the implemented ADRC-based system in MATLAB\slash Simulink is shown in Fig.~\ref{fig:example3}, and the results of using the proposed ADRC Toolbox in the above control problem are shown in Fig.~\ref{fig:results_converter}. The obtained results show that the applied ADRC function block can realise the given control objective by providing satisfactory results for output voltage tracking. Although the influence of the harmonic load resistance ($R_L$) has a visible effect on the shape of the control signal and the resultant output voltage, which is manifested through abrupt spikes of $\pm 0.2$V, the average value of the control error remained within a practically acceptable range. This example demonstrates that although ADRC is a powerful, robust control scheme, it has limitations. The challenging form of the external disturbance, on the one hand, invites the ADRC user to increase the observer and controller bandwidths to achieve better disturbance rejection and reference tracking, but on the other hand, it precludes obtaining high control performance owing to plant restrictions (e.g. finite actuation capabilities, sampling time) and ADRC function block limitations (e.g. the particular structure of the implemented observer; see~\cite{CEPconverter,CEPtorsionalFreq}). Hence, a compromise between the two should be found in engineering practice.

\subsection{Example \#4: Motor velocity control}
\label{sec:Example4}

\begin{figure}
    \centering
    \includegraphics[width=\textwidth]{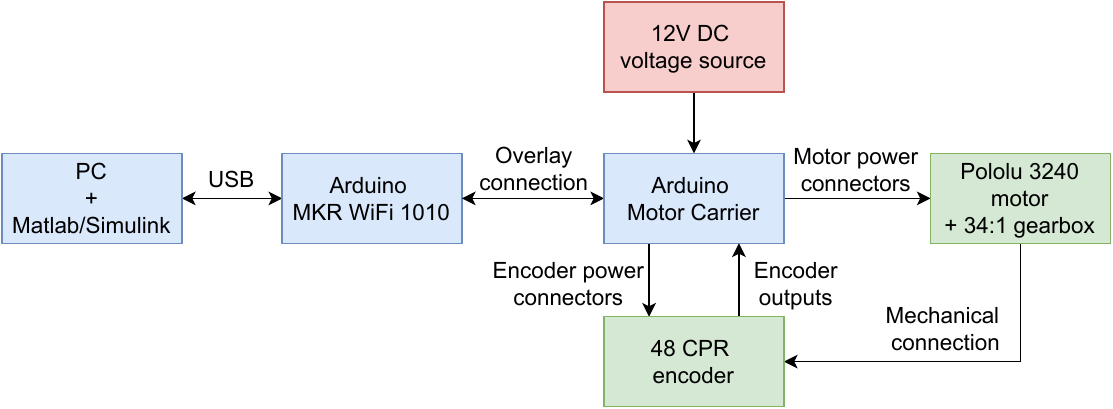}
    \caption{The experimental setup used in Example~\#4.}
    \label{fig:ExpSetupEx4}
\end{figure}

In this example, the controlled system is a Pololu 3240 gearmotor\footnote{\url{https://www.pololu.com/product/3240} (last visit: 20.09.2021)}, which consists of a 12V DC motor, 34:1 gearbox, and 48 CPR (counts per revolution) quadrature encoder integrated on the motor shaft. The system is governed by an Arduino MKR WiFi 1010 board with a dedicated Arduino MKR motor carrier with a sampling frequency of $100$Hz. The complete experimental setup is shown in Fig.~\ref{fig:ExpSetupEx4}. Such a low-cost, DIY-style configuration allows for rapid prototyping of various motion control applications. 

A simplified mathematical model describing the gearmotor dynamics can be formulated as a combination electrical and mechanical motor parts:
%
\begin{equation}
    \begin{cases}
        \frac{d^2\omega(t)}{dt^2} &= -\frac{R_aJ+L_ab_f}{L_aJ}\frac{d\omega(t)}{dt} - \frac{R_ab_f+k_\phi^2}{L_aJ}\omega(t) + \frac{k_\phi}{L_aJ}u(t) + d^*(t), \\
        y^*(t) &= \omega(t) + w(t), 
    \end{cases}
\end{equation}
where $u$[V] is the source voltage (control signal); $d^*$ is the external disturbance; $y^*$[rad/s] is the measured system output, which consists of the motor shaft angular velocity $\omega$[rad/s] and the sensor noise $w$[rad/s]; $R_a$[$\Omega$] is the armature resistance; $L_a$[H] is the armature inductance; $J$[kg$\cdot$m$^2$] is the rotor inertia; $b_f$[Nm/rad] is the friction coefficient; and $k_\phi$[Vs/rad] is the constant design parameter. Referring to the generalised form of the system description in \eqref{eq:originalDynamics}, we can assign $x(t):=\omega(t)$, $f^*(x,\dot{x},t) = -\frac{R_aJ+L_ab_f}{L_aJ}\frac{d\omega(t)}{dt} - \frac{R_ab_f+k_\phi^2}{L_aJ}\omega(t)$, and $b^* = \frac{k_\phi}{L_aJ}$.




The control objective here is to apply the control input $u$ that will make $\omega$ track a reference motor shaft angular velocity $x_d$[imp/sec] despite the presence of sensor noise and parametric uncertainty of the internal model dynamics.

\begin{figure}[t]
    \centering
    \includegraphics[width=1.0\textwidth]{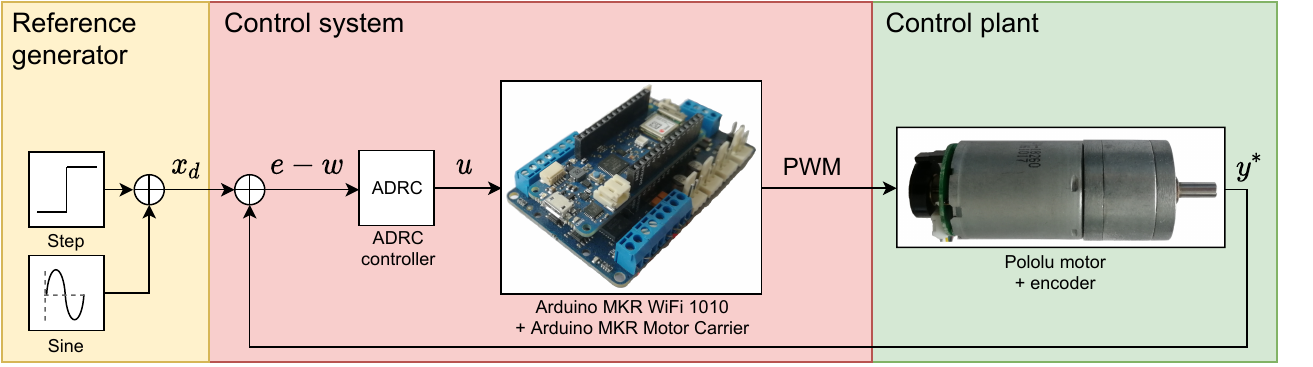}
    \caption{Block diagram of the DC motor control system (Example \#4) with the ADRC function block from the proposed ADRC Toolbox.}
    \label{fig:example4}
\end{figure}

\begin{figure}[t]
    \centering
    \includegraphics[width=0.32\textwidth]{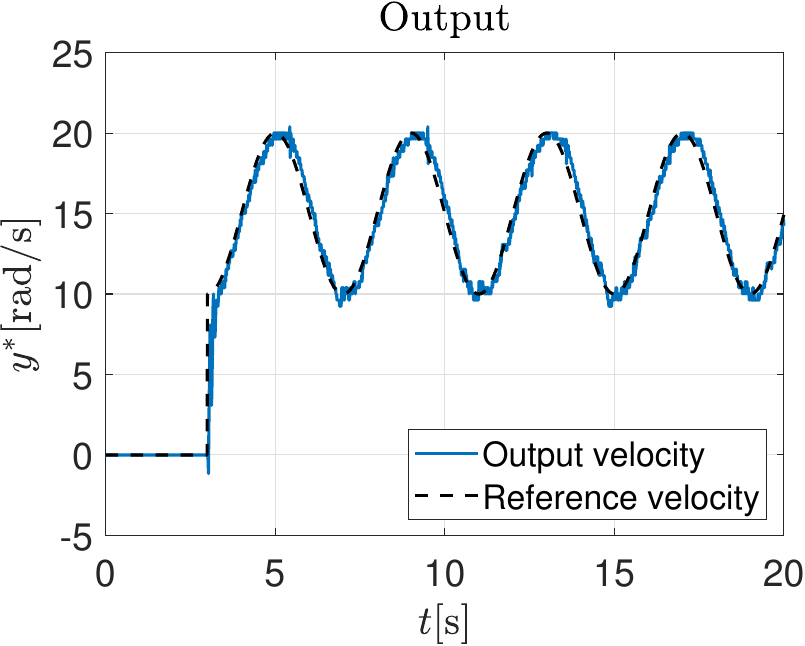}
    \includegraphics[width=0.32\textwidth]{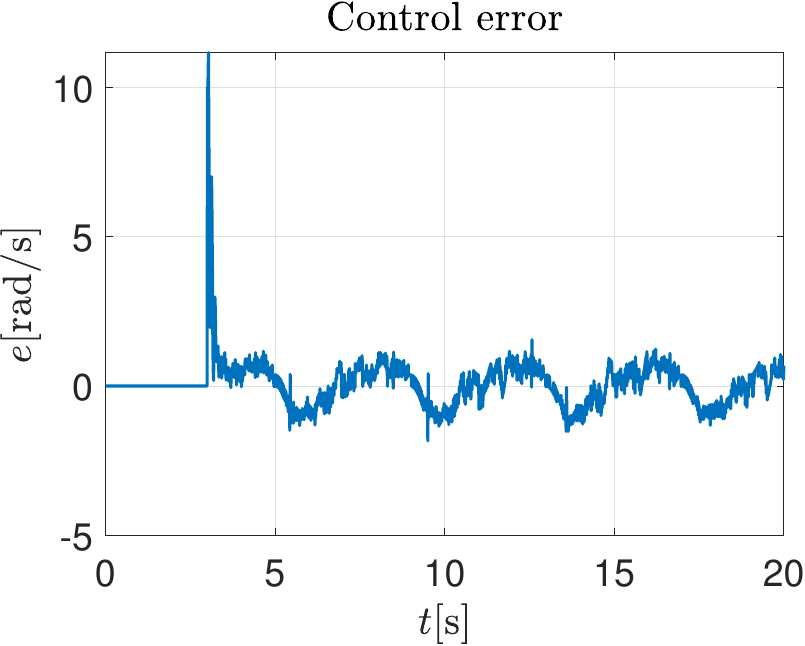}
    \includegraphics[width=0.32\textwidth]{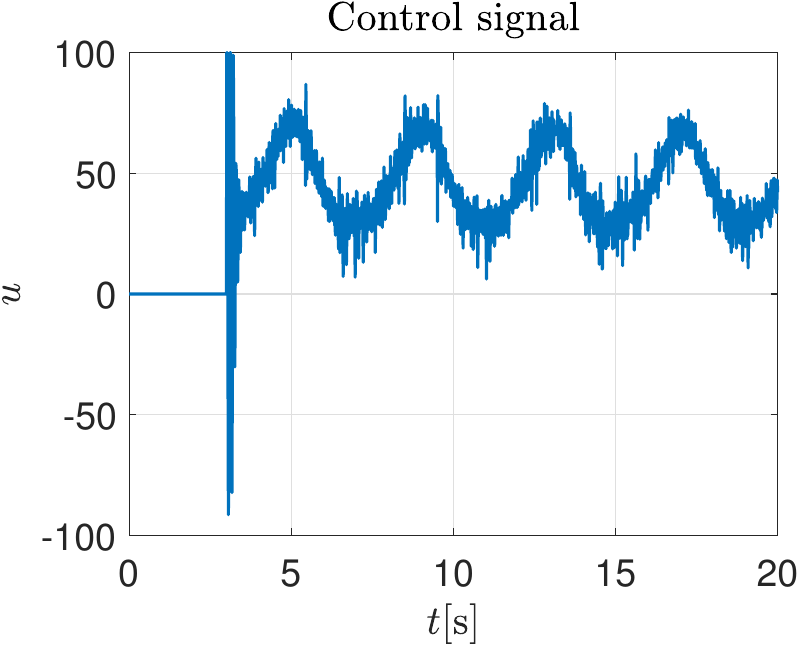}
    \caption{Experimental results of Example~\#4.}
    \label{fig:results_motor}
\end{figure}

In this example, the ADRC function block is implemented with the design parameters $n=2$, $\hat{b}=600$, $\omega_c=40$, and $\omega_o=90$, the anti-peaking mechanism turned off (default setting), and the saturation of the control signal turned on, with limits $u_{\text{min}}=-100$ and $u_{\text{max}}=100$. The block diagram of the implemented ADRC-based system in MATLAB\slash Simulink is shown in Fig.~\ref{fig:example4}, and the results of using the proposed ADRC Toolbox in the above control problem are shown in Fig.~\ref{fig:results_motor}. The obtained experimental results demonstrate the effectiveness of the ADRC function block in hardware applications. The motor output signal tracks the desired trajectory, which is a combination of a unit step and a harmonic function. As expected, both the control signal and the control error exhibited rapid changes when the system responded to the step change. Such an abrupt reaction, although kept at bay by the activated saturation mechanism, could be further minimised, for example, by adding a smoothing function, as was done in Example~\#3. Here, however, the step function was left deliberately to check how the control system performs in the case of an instantaneous change in the target signal (which often happens in engineering practice). It should be noted that the ADRC has not completely compensated for the harmonic component, which is present in both the control signal and the control error (inherited from the reference signal). Although the obtained control performance may be considered satisfactory, the implemented control structure causes the control error to oscillate around $\pm 1$rad/s. Here again, similar to Example~\#3, one could point to several specialised methods that would further increase the harmonic disturbance rejection. However, this is beyond the proposed ADRC Toolbox, at least in its current form.

\begin{remark}
\label{rem:Addons}
TO run and repeat the results from this hardware example, two MATLAB add-ons need to be installed: ''MATLAB Support Package for Arduino Hardware'' and ''Simulink Support Package for Arduino Hardware''.
\end{remark}

\subsection{Example \#5: Temperature control}
\label{sec:Example5}

\begin{figure}
    \centering
    \includegraphics[width=.45\textwidth]{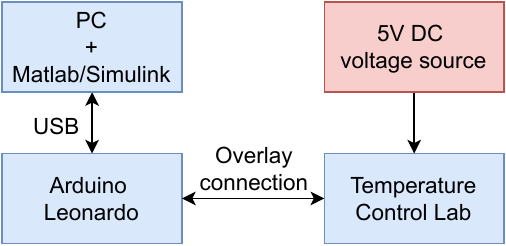}
    \caption{The experimental setup used in Example~\#5.}
    \label{fig:ExpSetupEx5}
\end{figure}

The system considered here is the ''Temperature Control Lab'' (TCLab)\footnote{\url{http://apmonitor.com/heat.htm} (last visit: 20.09.2021)}, which is a commercially available, portable, low-cost, Arduino-based temperature control kit~\cite{TClabIFAC}. The TCLab consists of two heaters and two temperature sensors. The experimental setup is shown in Fig.~\ref{fig:ExpSetupEx5}. In this specific example, a configuration with two operating heaters is used, which constitutes a MIMO control structure. The two heater units are placed in close proximity to each other to transfer heat by convection and thermal radiation. The system is governed with a sampling frequency of $10$Hz.

Based on~\cite{TCLabModel}, such a system can be described with the following nonlinear mathematical model, developed based on energy balance equations that represent the dynamics between the input power to each transistor and the temperature sensed by each thermistor.
\begin{equation}
    \begin{cases}
        \frac{dT_1(t)}{dt} &= f_1^*(T_1, T_2, t) + b_1^*u_1+d_1^*(t),\\
        y_1^*(t) &= T_1(t) + w_1(t),\\ 
        \frac{dT_2(t)}{dt} &= f_2^*(T_1, T_2, t) + b_2^*u_2+d_2^*(t),\\
        y_2^*(t) &= T_2(t) + w_2(t),\\
    \end{cases}
    \label{eq:TCLabModelTITO}
\end{equation}
where $u_1$[W] and $u_2$[W] are the output energy (control signals) from the first and second heaters, respectively; $y_1^*$[\textcelsius] and $y_2^*$[\textcelsius] are the measured system outputs, which consist of the first and second heater temperatures $T_1$[\textcelsius] and $T_2$[\textcelsius] and corresponding sensor noises $w_1$[\textcelsius] and $w_2$[\textcelsius]; $d_1^*$ and $d_2^*$ are the external disturbances that affect particular channels; and the functions $f_i^*(
\cdot) = \frac{1}{mC_p}\left[UA(T_{\infty}(t) - T_i(t)) + \epsilon\sigma A(T_{\infty}^4(t) - T_i^4(t)) + Q_{12}(t)\right]$ and $b_i^* = \frac{\alpha_i}{mC_p}$ for $i\in\{1,2\}$. In addition, $C_p$[J/kg-K] is the heat capacity, $T_\infty$[\textcelsius] is the ambient temperature, $U$[W/m$^2$-K] is the heat transfer coefficient, $A$[m$^2$] is the surface area not between the heaters, $A_s$[m$^2$] is the surface area between the heaters, $\epsilon$ is the emissivity, $m$[kg] is the mass, $\sigma$[W/m$^2$-K$^4$] is the Boltzmann constant, and $\alpha_i$[W/(\% heater)] is the $i$-th heater factor. The heat transfer exchange $Q_{12}$ between the heaters is a combination of convective heat transfer $Q_{C12} = UA_s(T_2 - T_1)$ and radiative heat transfer $Q_{R12} = \epsilon \sigma A(T_2^4 - T_1^4)$, defined as $Q_{12} = Q_{C12} + Q_{R12}$. A detailed derivation of the structure and parameters of~\eqref{eq:TCLabModelTITO} using physics-based and data-driven techniques can be found in~\cite{TCLabModel}. 

Referring to the generalised form of the system description in~\eqref{eq:originalDynamics}, one can assign $x(t):=T_1(t)$, $f^*(x,t) = \frac{1}{mC_p}\left[\right.UA(T_{\infty}(t) - T_1(t)) + \epsilon\sigma A(T_{\infty}^4(t) - T_1^4(t)) + Q_{12}(t)\left.\right]$, and $b^* = \frac{\alpha_1}{mC_p}$ for the first heater dynamics and $x(t):=T_2(t)$, $f^*(x,t) = \frac{1}{mC_p}\left[\right.UA(T_{\infty}(t) - T_2(t)) + \epsilon\sigma A(T_{\infty}^4(t) - T_2^4(t)) - Q_{12}(t)\left.\right]$, and $b^* = \frac{\alpha_2}{mC_p}$ for the second heater dynamics.

In this example, the control objective is to adjust both heater power outputs $u_{1/2}$ and transfer the thermal energy from the heaters to the temperature sensors to ensure that $T_{1/2}$ maintains the desired temperature set points $x_{d1/d2}$, despite the parametric uncertainty of the system model and sensor noise.  

\begin{figure}[t]
    \centering
    \includegraphics[width=0.75\textwidth]{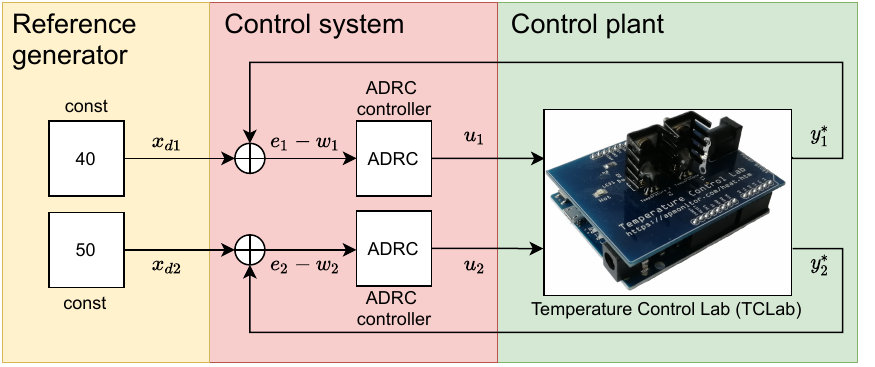}
    \caption{Block diagram of the temperature control lab control system (Example \#5) with the ADRC function blocks (from the proposed ADRC Toolbox), each of which controls one degree of freedom.}
    \label{fig:example5}
\end{figure}

The block diagram of the implemented ADRC-based system in MATLAB\slash Simulink is shown in Fig.~\ref{fig:example5}. One can see that in the performed example, the heaters are treated as two independent systems. The equations~\eqref{eq:originalDynamics} are therefore derived independently for each degree of freedom of the system~\eqref{eq:TCLabModelTITO}. Consequently, two ADRC function blocks are used, each to govern one input-to-output channel of the system. In such a configuration, the unmodelled influence of the cross-couplings is treated as part of the total disturbance of the system. 

\begin{figure}[t]
    \centering
        \includegraphics[width=0.32\textwidth]{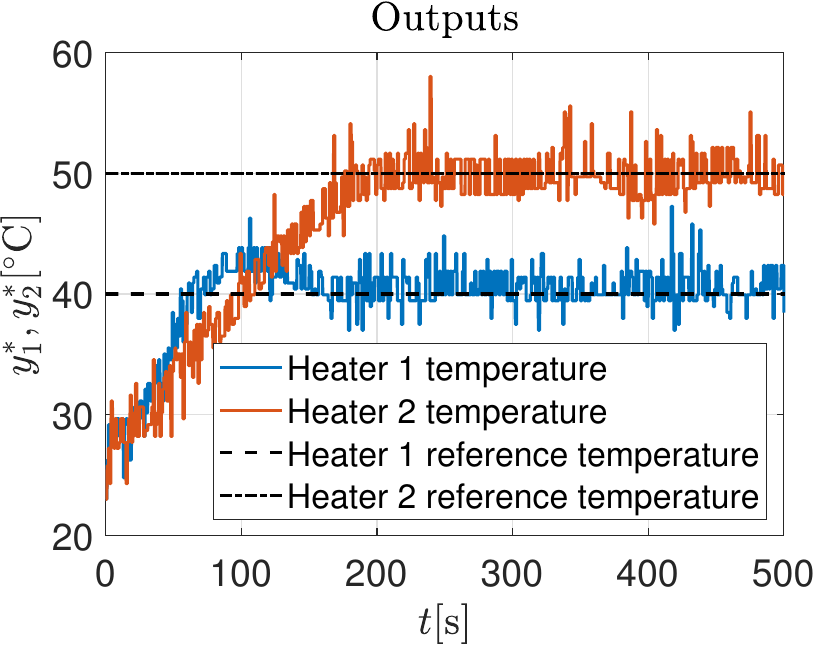}
        \includegraphics[width=0.32\textwidth]{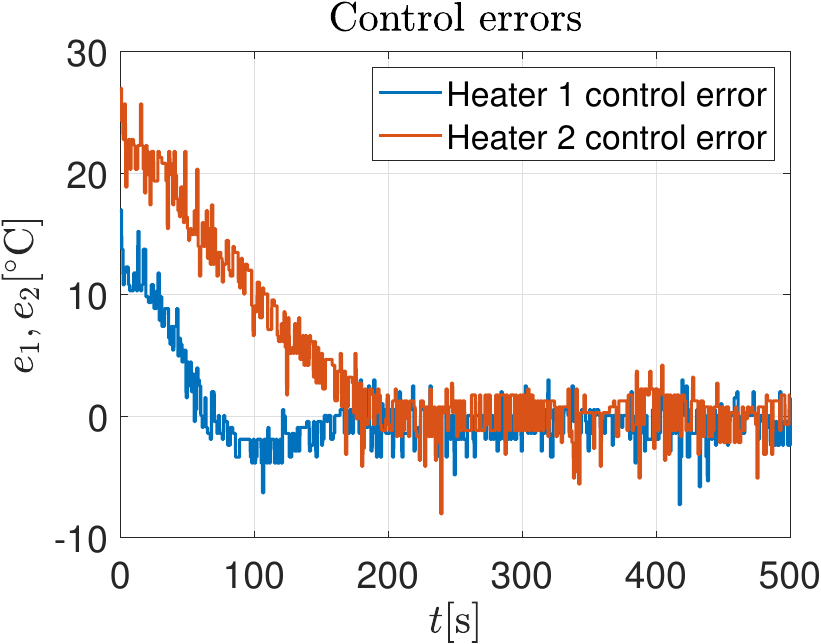}
        \includegraphics[width=0.32\textwidth]{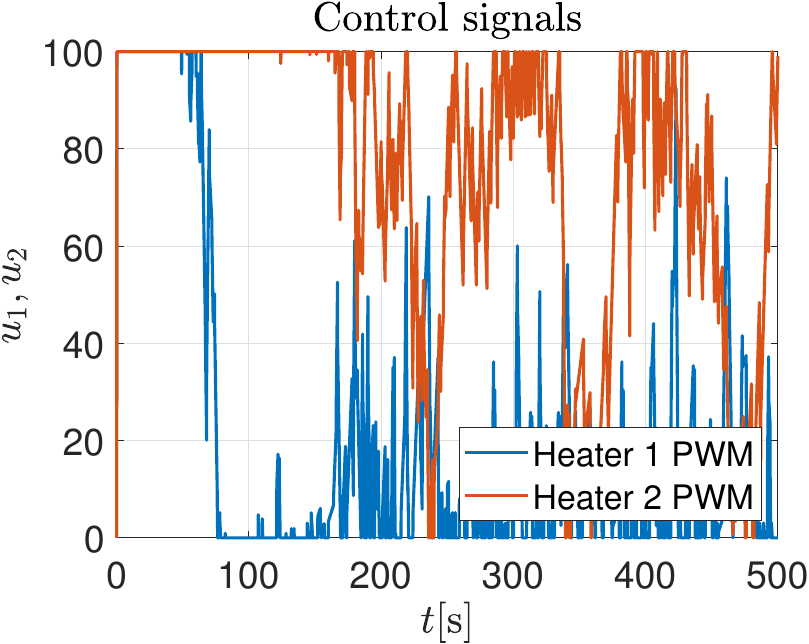}
    \caption{Experimental results of Example \#5.}
    \label{fig:results_tclab}
\end{figure}

The two ADRC function blocks are implemented here with design parameters $n=1$, $\hat{b}=3$, $\omega_c=6$, and $\omega_o=15$ for the first heater, and $n=1$, $\hat{b}=5$, $\omega_c=6$, $\omega_o=15$ for the second heater. Both function blocks have the anti-peaking mechanisms turned off (default setting), and the saturation of the control signals turned on, with user-defined limits $u_{\text{min}}=0$ and $u_{\text{max}}=100$. The results of using the proposed ADRC Toolbox to the above control problem are shown in Fig.~\ref{fig:results_tclab}. The experimental results validate the control structure with multiple ADRC function blocks as a viable solution for the considered cross-coupled MIMO plant. For the chosen tuning parameters, the target temperature set points were tracked with acceptable margins of error. Without the necessity of precise system modelling, the implemented robust ADRC approach managed to deal with plant modelling uncertainties, including the cross-couplings between the system's degrees of freedom. By comparing the selected ADRC parameters, one can see that a smaller estimated input gain ($\hat{b}$) was selected for the first heater. As a result, a steeper slope of convergence to the desired values was observer for the first heater output, which also resulted in a signal overshoot of approximately $3$\textcelsius. This is an expected result, which stems from the fact that $\hat{b}$ is also a scaling factor for the entire control signal (see~\eqref{eq:controller} and discussion thereof in Sect.~\ref{sec:Inputgain}). Regarding the control effort, one can see that the user-defined reference temperatures became challenging for the governed plant and pushed the control signals to extremes. This justifies the incorporation of the saturation mechanism in the proposed ADRC function block and shows the importance of activating saturation if the admissible range of the control signal is known. This is especially important in hardware experiments, where the control signals usually operate within certain physically limited bounds. It is also important for the considered observer-based control algorithm, which works correctly only if it is given accurate information about the input-output behaviour of the system (for details see Sect.~\ref{sec:Outputsaturation}).

\begin{remark}
\label{rem:Addons2}
To run and repeat the results from this hardware example, extra MATLAB add-ons need to be installed (see Remark~\ref{rem:Addons}) as well as dedicated TCLab libraries (available under the MIT licence at the producer website, mentioned earlier).
\end{remark}


\section{Software development and licensing}
\label{sec:Requirements}

The presented ADRC algorithm (Sect.~\ref{sec:ADRCpreliminaries}), toolbox parameters and functionalities (Sect.~\ref{sec:Overview}), and validation tests (Sect.~\ref{sec:Validation}) shown in this paper are for ADRC Toolbox version 1.1.3, which was current at the time of this paper's publication. The latest version of the ADRC Toolbox can be downloaded from \url{https://www.mathworks.com/matlabcentral/fileexchange/102249-active-disturbance-rejection-control-adrc-toolbox} and installed in MATLAB\slash Simulink by following the instructions therein. In addition to the necessary installation file, the website also contains contact information for the ADRC Toolbox developers and Q\&A section. The ADRC Toolbox can also be installed as an add-on directly from MATLAB using the ''Add-On Explorer'' function.

Regarding software development of the ADRC Toolbox, all related files, including MATLAB code, data, apps, examples, and documentation, are packaged in a single installation file (.mltbx), which greatly simplifies the sharing and installation of the ADRC Toolbox. This form of software development is convenient for end users, who can install the ADRC Toolbox without being concerned with the MATLAB path or other installation details, because the provided .mltbx file manages these details. 


\section{Conclusions}
\label{sec:Conclusions}

This work demonstrated the effectiveness and ease of use of the developed ADRC Toolbox for numerical simulations and hardware experiments performed in the MATLAB\slash Simulink software environment. The ADRC Toolbox is proposed to address the recent interest in the ADRC methodology and the growing use of MATLAB\slash Simulink in both academia and industry. Through a variety of case studies, it has been shown that the developed ADRC function block can be used for various applications and may be easily integrated with existing Simulink models. The implemented functionalities can help create custom ADRC applications swiftly and in a straightforward manner, thus potentially decreasing the time and effort usually required. The improvements on the implementation side of ADRC, introduced through the development of the ADRC Toolbox, are expected to make ADRC an even more compelling choice for solving real-world control problems.

The developed ADRC Toolbox is available to end users as open-source software. It is planned to be further developed with new functionalities based on feedback from the control community. Importantly, the ADRC Toolbox is an open-source project that is freely available to interested users. The current version of the ADRC Toolbox can be downloaded from \url{https://www.mathworks.com/matlabcentral/fileexchange/102249-active-disturbance-rejection-control-adrc-toolbox}. 

\section*{Acknowledgements}
\color{black}
\begin{itemize}
    \item The hardware setups used in the conducted experimental tests were purchased by R. Madonski and K. Łakomy with their private funds.
    \item R. Madonski was financially supported by the Fundamental Research Funds for the Central Universities project no. 21620335.
    \item W. Giernacki and J. Michalski were financially supported as a statutory work of the Poznan University of Technology, Poland (no. 0214/SBAD/0229).
    \item The example 'tclab\_temp\_control', provided alongside the ADRC Toolbox, uses a Simulink function block called 'Temperature Control Lab' as well as the 'TCLab' hardware platform, both available at \url{http://apmonitor.com/heat.htm}
\end{itemize}
\color{black}

\color{black}
\newpage
\section*{Appendix}
\label{sec:Appendix}

\begin{table}[h!]
\centering
\caption{\textcolor{black}{Parameters of systems used in Examples \#2, \#3, and \#5.}}
\label{tab:Parameters}
\begin{tabular}{|c|c|c|c|}
\hline
\multicolumn{4}{|c|}{\textbf{Example \#2}} \\ \hline
\textbf{Symbol} & \textbf{Unit} & \textbf{Value} & \textbf{Name} \\ \hline
$a$ & [m$^2$] & $7.5\cdot 10^{-5}$ & pipes cross-section area \\ \hline
$c$ & [m$^2$] & $1.2\cdot 10^{-2}$ & tanks cross-section area \\ \hline
$g$ & [m/s$^2$] & 9.81 & gravitational acceleration \\ \hline
\multicolumn{4}{|c|}{\textbf{Example \#3}} \\ \hline
\textbf{Symbol} & \textbf{Unit} & \textbf{Value} & \textbf{Name} \\ \hline
$V_{\text{in}}$ & [V] & 20 & input voltage source \\ \hline
$L$ & [H] & 0.01 & filter inductance \\ \hline
$C$ & [F] & 0.001 & filter capacitance \\ \hline
$R$ & [$\Omega$] & 50 & load resistance \\ \hline
\multicolumn{4}{|c|}{\textbf{Example \#5}} \\ \hline
\textbf{Symbol} & \textbf{Unit} & \textbf{Value} & \textbf{Name} \\ \hline
$\alpha_1$ & [W] & 0.01 & heater 1 factor \\ \hline
$\alpha_2$ & [W] & 0.01 & heater 2 factor \\ \hline
$C_p$ & [J/kg-K] & 500 & heat capacity \\ \hline
$A$ & [cm$^2$] & 10 & surface area not between heaters \\ \hline
$A_s$ & [cm$^2$] & 2 & surface area between heaters \\ \hline
$m$ & [kg] & 0.004 & mass \\ \hline
$U$ & [W/m$^2$-K] & 10 & overall heat transfer coefficient \\ \hline
$\epsilon$ & - & 0.9 & emissivity \\ \hline
$\sigma$ & [W/m$^2$-K$^4$] & $5.67\cdot 10^{-8}$ & Boltzmann constant \\ \hline
\end{tabular}
\end{table} 

\color{black}

\bibliography{mybibfile}

\end{document}